\documentclass{article}
\usepackage[T1]{fontenc}
\usepackage[utf8]{inputenc}
\usepackage[margin=1in]{geometry}
\usepackage{amsthm,amssymb,amsmath,mathtools,graphicx}
\usepackage{multirow,multicol,longtable,supertabular,tablefootnote,threeparttable}
\usepackage{wrapfig,lipsum,booktabs,tabularx,tabulary}
\usepackage{paralist,enumitem}
\usepackage{times}
\usepackage{soul}
\usepackage{fontawesome5}
\usepackage[switch]{lineno}
\usepackage[cal=cm]{mathalpha}
\usepackage{stix}
\usepackage{float}
\usepackage{algorithm}
\usepackage{algpseudocode}
\usepackage{verbatim}
\usepackage[round,semicolon]{natbib}
\usepackage{xcolor}
\usepackage{xspace}
\usepackage{textcomp}
\usepackage{makecell}
\usepackage{lscape} 
\usepackage{siunitx}
\usepackage{pifont}
\usepackage{scrextend}
\usepackage{array}
\usepackage{tgpagella}
\usepackage{url}         
\usepackage{nicefrac}       
\usepackage{changepage}
\usepackage{xargs}          
\usepackage{endnotes}
\usepackage{pgfplots}
\usetikzlibrary{pgfplots.groupplots}
\pgfplotsset{compat=1.3}
\usepackage{tikz}
\usetikzlibrary{patterns}
\usepackage[most]{tcolorbox}
\usepackage{minitoc}
\usepackage{xfrac}
\usepackage{setspace}

\definecolor{mydarkblue}{rgb}{0,0.08,0.45}
\usepackage[colorlinks,citecolor=mydarkblue,urlcolor=mydarkblue,linkcolor=mydarkblue]{hyperref}

\usepackage{subcaption}

\usepackage[export]{adjustbox}
\usepackage[capitalize,noabbrev]{cleveref}

\crefname{section}{Section}{\S\S}
\Crefname{section}{Section}{\S\S}
\crefname{table}{Table}{Tables}
\crefname{figure}{Figure}{Figures}
\crefname{algorithm}{Algorithm}{}
\crefname{equation}{eq.}{}
\crefname{appendix}{Appendix}{}
\crefformat{section}{Section #2#1#3}

\newcommand{\method}{\texttt{EncCluster}}
\makeatletter
\newcommand{\ssymbol}[1]{^{\@fnsymbol{#1}}}
\makeatother

\begin{document}
\date{}

\title{\textbf{\Large{EncCluster: Scalable Functional Encryption in Federated Learning through Weight Clustering and Probabilistic Filters}}}

\author{
\normalsize{}
$^{*}$Vasileios Tsouvalas$^1$ \qquad $^{*}$Samaneh Mohammadi$^{2,3}$ \qquad Ali Balador$^3$ \\
\normalsize{}
Tanir Ozcelebi$^1$ \qquad Francesco Flammini$^3$ \qquad Nirvana Meratnia$^1$ \\
\normalsize{}
$^1$Eindhoven University of Technology, The Netherlands \\
\normalsize{}
$^2$RISE Research Institutes of Sweden, Sweden \\
\normalsize{}
$^3$Mälardalen University, Sweden \\
}

\footnotetext[1]{Equal Contribution. Corresponding authors: v.tsouvalas@tue.nl${^1}$, samaneh.mohammadi@ri.se${^2}$}

\maketitle

\begin{abstract}
\noindent Federated Learning (FL) enables model training across decentralized devices by communicating solely local model updates to an aggregation server. Although such limited data sharing makes FL more secure than centralized approached, FL remains vulnerable to inference attacks during model update transmissions. Existing secure aggregation approaches rely on differential privacy or cryptographic schemes like Functional Encryption (FE) to safeguard individual client data. However, such strategies can reduce performance or introduce unacceptable computational and communication overheads on clients running on edge devices with limited resources. In this work, we present~\method\footnote[2]{Code will be release on \href{https://github.com/FederatedML/EncCluster}{GitHub}.}, a novel method that integrates model compression through weight clustering with recent decentralized FE and privacy-enhancing data encoding using probabilistic filters to deliver strong privacy guarantees in FL without affecting model performance or adding unnecessary burdens to clients. We performed a comprehensive evaluation, spanning various datasets and architectures, to demonstrate~\method~scalability across encryption levels. Our findings reveal that~\method~significantly reduces communication costs \textemdash~below even conventional FedAvg \textemdash~and accelerates encryption by more than four times over all baselines; at the same time, it maintains high model accuracy and enhanced privacy assurances.
\end{abstract}

\section{Introduction\label{sec:intro}}

The vast amount of data generated by smartphones, wearables, and Internet of Things devices has been instrumental in advancing Machine Learning (ML) systems. However, traditional approaches used to aggregate data in centralized servers for processing face increasing challenges due to privacy concerns and regulations such as the General Data Protection Regulation\footnote[3]{https://gdpr-info.eu/} and the Artificial Intelligence Act\footnote[4]{https://artificialintelligenceact.eu/}. In response, Federated Learning (FL) has emerged, enabling collaborative training of AI models directly on edge devices (referred to as clients), locally with on-device data~\cite{konevcny2016federated}. This process involves multiple federated rounds, which include dispatching model updates from the server to clients, conducting local training on clients, and finally aggregating these updates on the server side (e.g., \textit{FedAvg}~\cite{mcmahan2017communication}). This process is performed iteratively until the model reaches convergence. 

Although FL offers data privacy by keeping raw data on devices, it remains vulnerable to inference attacks during model update communications between clients and servers, which could expose sensitive client information, as has been demonstrated in recent studies~\cite{nasr2019comprehensive,shokri2017membership}. To address this concern, various privacy-preserving mechanisms for FL have been proposed~\cite{bonawitz2017practical,aono2017privacy,truex2019hybrid,wei2020federated} and techniques such as Differential Privacy (DP)\cite{abadi2016deep} and Secure Multi-Party Computation (SMPC)~\cite{bonawitz2017practical} have been employed. However, DP can impact model performance~\cite{yang2023privatefl}, while SMPC offers security advantages but struggles with scalability due to high computational demands~\cite{riazi2020heax}. 

As another alternative, recent encryption-based methods (see Table~\ref{tab:existingwork}), such as Homomorphic Encryption (HE)~\cite{fang2021privacy} and Functional Encryption (FE)~\cite{xu2019hybridalpha}, offer strong privacy guarantees with efficient cryptographic operations and simpler communication protocols (that is, a single round trip per federated round) without affecting model performance. These approaches often require a fully trusted third party for key management, creating a significant hurdle in scenarios where such trust is hard to establish or third parties are risk-prone. At the same time, encrypting the entire set of local model parameters incurs significant computational costs and extends FL training times. This poses challenges for edge devices, which have limited computational and energy resources, underlining the urgency for efficient FL encryption practices. Moreover, the computational cost of these encryption-based schemes tends to rise exponentially with increased security levels, while the communication overhead is linearly affected by the key size of encryption~\cite{mohammadi2023secure}. This underscores a fundamental trade-off in FL between ensuring privacy and maintaining an acceptable level of computational and communication overhead.

\begin{table*}[!t]
    \centering
    \caption{Summary of existing works on privacy-preserving methods in FL}\label{tab:existingwork}
    \resizebox{0.95\textwidth}{!}{%
        \begin{tabular}{lccccccc}
            \toprule
            \textbf{Approach} 
            & \multirow{2}{*}{\begin{tabular}[c]{@{}c@{}}\textbf{Privacy}\\\textbf{Mechanism}$^{\dagger}$\end{tabular}}
            & \multicolumn{3}{c}{\textbf{Threat Model}} 
            & \multicolumn{2}{c}{\textbf{Efficiency}} 
            & \multirow{2}{*}{\textbf{Model Fusion}} \\
            \cmidrule(lr){3-5}
            \cmidrule(lr){6-7}
            & & \textit{Clients} & \textit{Server} & \textit{TTPA$^{\ddagger}$} & \textit{Communication} & \textit{Computation} & \\
            \midrule
            BatchCrypt~\cite{zhang2020batchcrypt} & HE & Honest & HbC$^{\S}$ & Not-required & \checkmark & \checkmark & \textit{additive} \\
            CryptoFE~\cite{qian2022cryptofe} & FE & HbC$^{\S}$ & HbC$^{\S}$ & Required & & & \textit{additive} \\
            SEFL~\cite{mohammadi2023secure} & HE & Honest & HbC$^{\S}$ & Required & \checkmark & \checkmark & \textit{additive} \\
            HybridAlpha~\cite{xu2019hybridalpha} & FE+DP & Dishonest & HbC$^{\S}$ & Required & & & \textit{weighted} \\
            DeTrust-FL~\cite{xu2022detrust} & FE+DP & HbC$^{\S}$ & HbC$^{\S}$ & Not-required & & & \textit{weighted} \\
            \textbf{\method~(Ours)} & FE+BF & Dishonest & HbC$^{\S}$ & Not-required & \checkmark & \checkmark & \textit{weighted} \\
            \bottomrule
        \end{tabular}%
    }

    \begin{tablenotes}
        \scriptsize{
        \item $\dagger$ FE: Functional Encryption, DP: Differential Privacy, HE: Homomorphic Encryption, BF: Binary Fuse filters
        \item $\ddagger$ TTPA: Trusted Third Party Authority
        \item $\S$ HbC: Honest-but-Curious
        }
    \end{tablenotes}
    \vspace{-12pt}
\end{table*}

In this paper, we propose the~\method~framework, to offer robust privacy protection against inference attacks while requiring minimal communication and computation overhead for clients participating in FL. To achieve this, we design our framework with the following three building blocks, i.e. (i) model compression via weight clustering, (ii) decentralized FE, allowing cryptographic encryption without a fully trusted third party, and (iii) encoding via probabilistic data structure, termed Binary Fuse (BF) filters, to enhance privacy without introducing excessive computational burdens. Specifically, we apply weight clustering locally on clients' models and subsequently encrypt the resulting set of cluster centroids via FE. Cluster-weight mapping, which signifies associations between positions in the weight matrix and respective centroids, is then injected into BF filters through computationally efficient hashing operations. To fuse all model updates, the server reconstructs this mapping via a membership query in the BF filters and performs a secure aggregation without decrypting the clients' model updates. In doing so,~\method~restricts the computationally ``\textit{heavy}'' encryption operations to a small set of centroid values, while their mapping to model weights is encoded through cost-effective hashing operations, striking a balance between preserving privacy and meeting practical computational and communication demands in FL. Concisely, our contributions are:

\begin{itemize}[noitemsep]
    \vspace{-10pt}
    \item We present~\method, a framework that enhances FL privacy against inference attacks with minimal effect on model performance, while reducing communication and computation costs.

    \item We combine weight clustering with decentralized FE and BF filter-based encoding for secure, efficient transmission, and aggregation of compressed model updates, eliminating reliance on trusted external entities.

    \item Our comprehensive evaluation across diverse datasets and federated settings, demonstrates \texttt{EncClu} \texttt{ster}'s significant efficiency gains over traditional FE schemes, achieving $13$ times reduction in communication costs and $160$-fold increase in computational efficiency, alongside a mere 1.15\% accuracy loss compared to \textit{FedAvg}.

    \item We showcase~\method~scalability across various encryption levels and four network architectures, demonstrating near-constant communication costs in FL with minimal increases to clients' encryption times, all while maintaining robust privacy guarantees.
\end{itemize}

\section{Preliminaries\label{sec:prelim}}

\noindent \textbf{Federated Learning.} Federated Learning is a collaborative learning paradigm that aims to learn a single, global model from data stored on remote clients with no need to share their data with a central server. Specifically, with the data residing on clients' devices, a subset of clients is selected to perform a number of local \textit{SGD} steps on their data in parallel on each communication round. Upon completion, clients exchange their models' weight updates with the server, aiming to learn a unified global model by aggregating these updates. Formally, the goal of FL is typically to minimize the following objective function:

\vspace{-0.1cm}
\begin{equation} \label{eqn:FL}
     \min_{\theta} \mathcal{L}_{\theta} = \sum_{i=1}^{N} \nu_{i} \cdot {\mathcal{L}}_i(\theta)~,
\end{equation}

\noindent where $\mathcal{L}_i$ is the loss function of the $i$-th client and $\nu_{i}$ corresponds to the relative impact of the $i$-th client on the construction of the globally aggregated model. For the \textit{FedAvg}~\cite{mcmahan2017communication} algorithm, $\nu_{i}$ is equal to the ratio of client's local data $|D_{i}|$ over all training samples, i.e., $\left (\nu_{i} = \frac{|D_{i}|}{|D|}\right)$. 

\noindent \textbf{Weight Clustering.} Weight clustering is a neural network model compression technique in which similar weights are grouped into clusters using a clustering algorithm such as K-means~\cite{lloyd1982least}. This process can be executed either per layer, clustering each layer's weights independently, or across the entire model, clustering all weights collectively. Given a neural network $\textrm{f}$ with weight parameters $\theta = (\theta_1, \dots, \theta_d) \in \mathbb{R}^d$, the objective of the clustering algorithm is to identify $\kappa$ distinct clusters $\mathcal{C} = \{ c_1, \ldots, c_\kappa \}$, with the aim of minimizing the following objective function:

\vspace{-0.2cm}
\begin{equation}\label{eqn:cluster}
\mathcal{L}_{wc}(\theta, \mathcal{Z}) = \sum_{j=1}^{\kappa} \sum_{i=1}^{d} u_{ij} \cdot || \theta_i - z_j ||^2 ~,
\end{equation}

\noindent where $\mathcal{Z} = \{z_1, \dots, z_\kappa \}$ represents the set of $\kappa$ centroids, each corresponding to a cluster $c_i$. The term $||\cdot||$ denotes the Euclidean distance operator, and $u_{ij}$ is a binary indicator function that returns $1$ when weight $\theta_i$ belongs to cluster $c_j$ and $0$ otherwise. In essence, $\mathcal{L}_{wc}$ calculates the sum of squared Euclidean distances between each weight and its nearest centroid, weighted by $u_{ij}$. Upon minimizing $\mathcal{L}_{wc}$, we obtain the set of centroids $\mathcal{Z}$ and their cluster position matrix $\mathcal{P}$ (referred to as cluster-weights mappings), mapping each point of $\theta$ to its corresponding cluster centroid value in $\mathcal{Z}$. 

\noindent \textbf{Probabilistic Filters.} Probabilistic filters are specific data structures that map a universe of keys, denoted as $\mathcal{U}$, of varying bit lengths, to fixed-size bit values, thereby compacting real-world data representations effectively. They achieve this by using hash functions to transform and store data in a uniformly distributed array, known as the fingerprints $\mathcal{H}$. This compact representation $\mathcal{H}$ facilitates efficient membership checking, with an adjustable rate of false positives \textemdash~where a non-member might be incorrectly identified as a member \textemdash~while ensuring zero false negatives. While there are multiple variations of probabilistic filters, we focus on \textit{Binary Fuse} (BF) filters~\cite{graf2022binary}, which are known for their exceptional space efficiency and computational effectiveness. These filters achieve significant space efficiency, up to $1.08$ times the bits allocated per entry (bpe), and maintain low false positive rates given by $2^{-bpe}$, enabling scalability to desired precision in both efficiency and false positive rate.

Formally, a $\mu$-wise BF utilizes $\mu$ distinct hash functions $h_{j}$:$~\{1,\ldots,2^n\} \rightarrow \{1,\ldots, t\}$ for $j \in \{1,\ldots,\mu\}$, where $t$ denotes the size of the fingerprints array, $\mathcal{H}$. Let $g$:$~\mathbb{N} \rightarrow \{1,\ldots,2^n\}$ be the fingerprint generation function, mapping each key to an $\xi$-bit value. For a given set of keys $\mathcal{U}$, we can compute the fingerprint array $\mathcal{H}$ as:

\begin{equation}\label{eq:filter_in}
    \begin{aligned}
        \mathcal{H} = \bigcup_{i \in \mathcal{U}} \phi(i) = \bigcup_{i \in \mathcal{U}} \left ( \bigcup_{j=1}^{\mu} \{ h_{j}(g(i)) \} \right )
    \end{aligned}
\end{equation}
\vspace{10pt}

Here, $\phi(i)$ computes the set of $\mu$ locations in $\mathcal{H}$ for each key $i$ in $\mathcal{U}$. Once $\mathcal{H}$ is constructed, we can perform a membership check as:

\vspace{-0.2cm}
\begin{equation}\label{eq:filter_check}
    \mathrm{Member}(x) = 
    \begin{cases}
        \mathrm{true,} &  \bigoplus_{j=1}^{m} \mathcal{H} \left [ h_j(g(x)) \right ] = g(x) \\
        \mathrm{false,} & \mathrm{otherwise}
    \end{cases}
\end{equation}

\noindent where, $\bigoplus_{j=1}^{m} \mathcal{H}[\cdot]$ represents the bitwise \textit{XOR} operation performed on the array values of $\mathcal{H}$, indicated by the hash functions $h_j(g(x))$. The $\mathrm{Member}$$(\cdot)$ function returns true if the result of the \textit{XOR} operation over $\mathcal{H}$ matches the fingerprint $g(x)$, suggesting that $x$ is likely a member of the set, and returns false in all other occasions. It is important to note that while computing a large number of hashes may seem daunting, not all hashing algorithms are computationally expensive. For example, BF filters use MurmurHash3~\cite{appleby2016murmurhash3}, which is computationally efficient and exhibits exceptional properties for hashing large data structures into space-efficient arrays (e.g., uniform hash distribution and randomness). 

\noindent \textbf{Decentralized Functional Encryption.} Functional encryption is a cryptographic paradigm that enables operations over encrypted data to yield plaintext results without decrypting individual inputs~\cite{boneh2011functional}. Compared with traditional HE methods, FE demonstrates significant efficiency improvements, particularly in secure multi-party aggregation tasks~\cite{xu2019hybridalpha}. In this paper, we focus on Decentralized Multi-Client Functional Encryption (DMCFE)~\cite{chotard2018decentralized}, a variant of FE that facilitates inner product computations on encrypted data and is distinguished by two key features, i.e., (i)  allowing each participant to hold a partial key, thereby obviating the need for a trusted Third Party Authority (TPA) to generate functional keys during the FL process, and (ii) introducing a labeling mechanism that cryptographically binds a specific functional key to a specific ciphertext, ensuring each key's exclusivity to its designated ciphertext.

Let $\mathcal{F}$ be a family of sets of functions $f$:$~\mathcal{X}_{1} \times \ldots \times \mathcal{X}_{n} \to \mathcal{Y}$, $\ell = \{0, 1\}^{*} \cup \{\bot\}$ be a set of labels, and $\mathcal{N}$ be a set of clients. A DMCFE scheme for the function family $\mathcal{F}$ and the label set $\ell$ is a tuple of six algorithms $\mathcal{E}_{DMCFE} = (\mathsf{Setup}, \mathsf{KeyGen}, \mathsf{dKeyShare},\mathsf{dKeyComb}, \mathsf{Enc}, \mathsf{Dec})$:

\begin{itemize}
    \item $\mathsf{Setup}(\lambda, n)$: Takes as input a security parameter $\lambda$ and the number of clients $n$ and generates public parameters $\mathsf{pp}$. We will assume that all the remaining algorithms implicitly contain $\mathsf{pp}$.

    \item $\mathsf{KeyGen}(\mathsf{id}_i)$: Takes as input a client-specific identifier, $\mathsf{id}_i$, and outputs a secret key $\mathsf{sk}_i$ and an encryption key $\mathsf{ek}_i$, unique to client $i$.

    \item $\mathsf{dKeyShare}(\mathsf{sk}_i, f)$: Takes as input a secret key $\mathsf{sk}_i$ and a function $f \in \mathcal{F}$ to computes a partial functional decryption key $\mathsf{dk}_i$.

    \item $\mathsf{dKeyComb}(\{\mathsf{dk}_i\}_{i \in \mathcal{N}})$: Takes as input a set of $n$ partial functional decryption keys $\{\mathsf{dk}_i\}_{i \in \mathcal{N}}$ and outputs the functional decryption key $\mathsf{dk}_f$.

    \item $\mathsf{Enc}(\mathsf{ek}_i, x_{i,l})$: Takes as input an encryption key $\mathsf{ek}_i$ and a message $x_i$ to encrypt under label $l \in \ell$ and outputs ciphertext \( \mathsf{ct}_{i,l} \).

    \item $\mathsf{Dec}(\mathsf{dk}_f, \{\mathsf{ct}_{i,l}\}_{i \in \mathcal{N}})$: Takes as input a functional decryption key $\mathsf{dk}_f$, and $n$ ciphertexts under the same label $l$, and computes value $y \in \mathcal{Y}$.
\end{itemize}

\begin{figure*}[!t]
    \centering
    \resizebox{0.95\textwidth}{!}{
        \includegraphics[width=\linewidth]{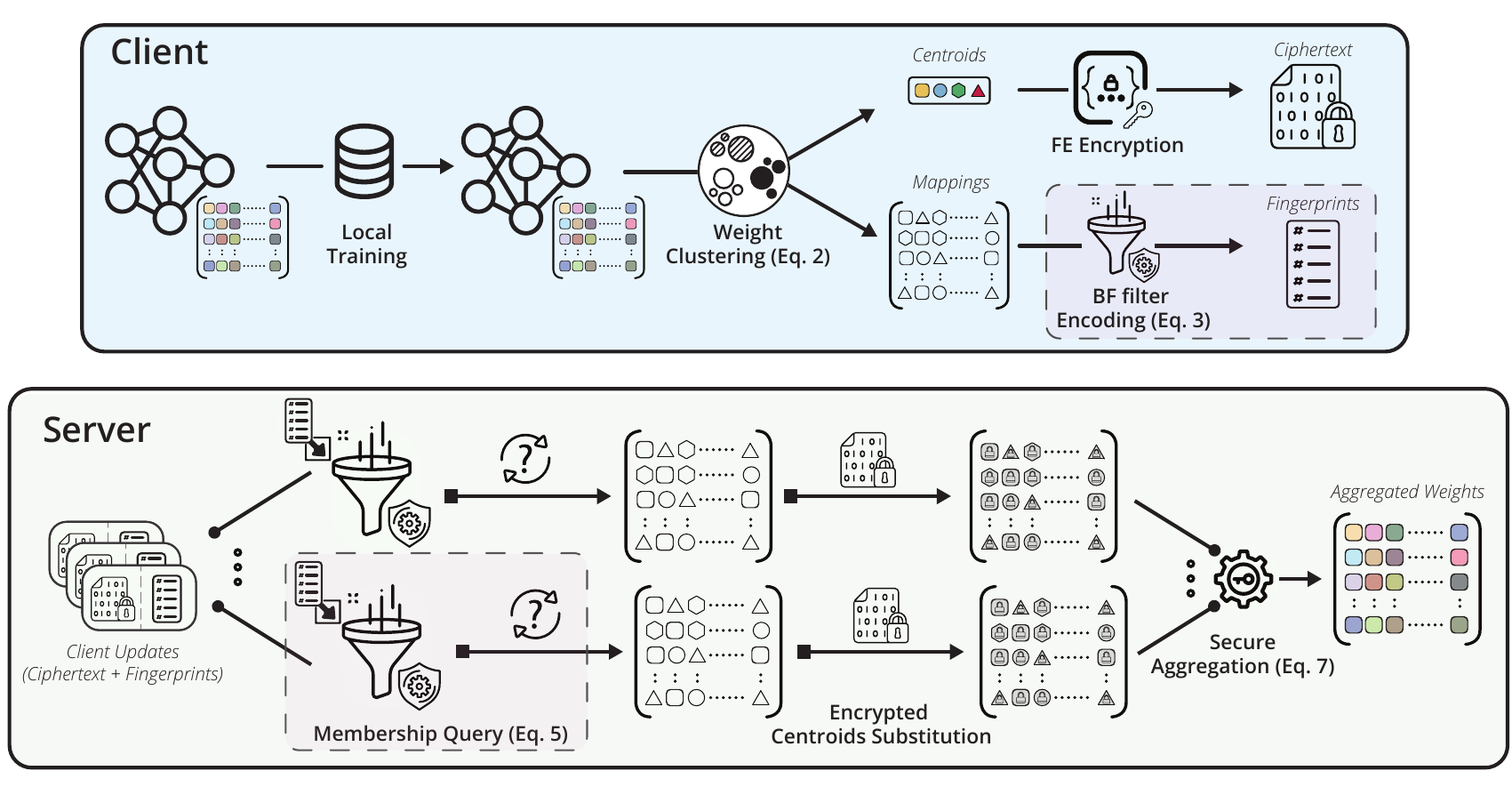}
    }
    \caption{Overview of~\method's training process: Clients train on local data, cluster weights, then encrypt \textit{centroids} using DMCFE into \textit{ciphertexts} and encode cluster-weight $mappings$ into BF filter \textit{fingerprints} (hashed array). The server reconstructs $mappings$ via BF filter queries using \textit{fingerprints}, derives encrypted clustered weights, and aggregates them to update the model.
    } \label{fig:overview}
\end{figure*}

\section{\texttt{EncCluster} Framework\label{sec:method}}

We present~\method, a novel approach that combines model compression via weight clustering, decentralized FE, and BF filter-based efficient encoding to enhance FL's privacy against inference attacks, while simultaneously reducing communication costs and the computational load on clients.

\subsection{Notations}
We use the following notation in the rest of the paper. Let $\mathcal{A}$ denote the aggregation server, and $\mathcal{N}$ represent the set of $n$ clients, each with their local dataset $\mathcal{D}n$, participating over $R$ federated rounds in the FL process. The neural network model is expressed as $\textrm{f}{\theta}$, characterized by weight parameters $\theta = (\theta_1, \dots, \theta_d) \in \mathbb{R}^d$. The notation $\theta^*$ refers to the optimized parameters after local training. For a given client $n$ and round $r$, $\mathcal{C}_n^r$ denotes the set of $\kappa$ clusters formed from the client's private model weights $\theta_n^{r}$, where $\mathcal{Z}_n^r$ represents the cluster centroids, and $\mathcal{P}_n^r$ maps each weight in $\theta$ to its corresponding cluster centroid in $\mathcal{Z}_n^r$. Additionally, $\mathcal{H}_n^r$ indicates the fingerprint array (hashed cluster-weights mappings) for client $n$ in round $r$, while $\hat{x}$ refers to the encrypted version of any value $x$ using the DMCFE~\cite{chotard2018decentralized}. Note that we use the current round $r$ as label $l$ in the DMCFE, similar to~\cite{xu2019hybridalpha,xu2022detrust,mohammadi2023optimized}, which we omit for simplicity. We provide the full list of abbreviations along with their definitions in Table~\ref{tab:notations}.

\subsection{Threat Model and Assumptions}

Our threat model has $3$ independent entities:

\begin{itemize}
    \setlength\itemsep{0em}
    \item \textbf{\textit{HbC Server}}: Assumes adherence to algorithms and protocols, with an interest in learning private information from specific clients' model updates, potentially colluding with other clients.
    \item \textbf{\textit{Dishonest clients}}: Participants who may collude with the server to access other clients' private information, without manipulating or poisoning model updates.
    \item \textbf{\textit{HbC TPA}}: An independent entity that follows the DMCFE protocol but may seek private information from specific clients.
\end{itemize}

\noindent As the DMCFE protocol prevents immediate access to clients' model weights, the HbC server cannot directly seek private information; however, it might still perform malicious activities (e.g., selectively aggregating updates) to infer clients' updates. Consequently, our privacy analysis extends to targeted active attacks initiated by the server, including isolation and replay attacks, with or without client collusion. Dishonest clients can only collude by providing the server with their model weights and cannot request other clients' private information. Our framework relies on an HbC TPA throughout the FL process to initiate the DMCFE protocol as needed. 

To mitigate risks like man-in-the-middle attacks and snooping, all communications occur over secure channels. Key confidentiality is ensured through a secure key-provisioning procedure, such as Diffie-Hellman \cite{canetti2001analysis}. Note that our analysis does not cover denial-of-service or poisoning attacks, as these threats are beyond the scope of this work. Aligned with our threat model and operational assumptions~\method~ensures: (i) the \textit{HbC server} only accesses the aggregated global model, (ii) \textit{dishonest clients} cannot access honest clients' updates, and (iii) the \textit{HbC TPA} remains unaware of clients' details.

\subsection{EncCluster Overview}

\begin{minipage}{0.6\textwidth}
\noindent \textbf{DMCFE Initialization.}~\method~initiates with a TPA executing $\mathsf{Setup}(\lambda, n)$ to generate public parameters $\mathsf{pp}$, which all parties can access. With a unique identifier $\mathsf{id}_i$ from the TPA, each client independently creates their secret and encryption key pairs $(\mathsf{sk}_i, \mathsf{ek}_i)$ using $\mathsf{pp}$. By doing this, clients are able to generate partial decryption keys using $\mathsf{dKeyShare}(\mathsf{sk}_i)$. In every round, clients send their partial decryption keys along with their model updates back to the server, facilitating the ongoing FL training process. \\

\noindent \textbf{Training Process}: The~\method~training pipeline begins with the server randomly initializing a neural network $\textrm{f}_{\theta}$, which distributes to all clients. The clients then encrypt $\mathcal{Z}^r_n$ using their encryption key  $\mathsf{ek_n}$ to derive the encrypted centroids set, $\hat{\mathcal{Z}}^r_n$, and insert the entries of $\mathcal{P}^r_n$ in a $\mu$-wise BF filter, resulting in fingerprint array $\mathcal{H}^r_n$. The pair of $\hat{\mathcal{Z}}^r_n$  and $\mathcal{H}^r_n$  is communicate to server,
\end{minipage} \qquad  
\begin{minipage}{0.35\textwidth}
    \centering \small
    \captionof{table}{\small{Key notations in~\method.}} \label{tab:notations}
    \resizebox{1.0\linewidth}{!}{
        \begin{tabular}{cl}
            \toprule
            \textbf{Notation} & \multicolumn{1}{c}{\textbf{Definition}} \\
            \midrule
            $\mathcal{A}$ & Aggregation server \\
            $\mathcal{N}$ & Set of $N$ clients \\
            $\mathcal{D}_n$ & Dataset held by $n$-th client \\
            $\textrm{f}_{\theta}$ & Neural network with parameters $\theta$ \\
            $\theta^*$ & Post-training model parameters \\
            $\mathcal{L}_{ce}$ & Cross-Entropy Loss \\
            $\mathcal{L}_{wc}$ & Weight Clustering Loss \\
            $\kappa$ & Number of clusters \\
            $\mathcal{C}$ & Set of $\kappa$ clusters \\
            $\mathcal{Z}$ & Clusters' centroids \\
            $\mathcal{P}$ & Cluster-weights mapping \\
            $\mathcal{H}$ & Fingerprint (hashed) array of BF filter \\
            \cmidrule(lr){1-2}
            $\hat{x}$ & Encrypted vector of $x$ using DMCFE \\
            $|x|$ & Number of elements in $x$ \\
            $\{x\}$ & Set containing the values from 0 to $x$ \\
            \bottomrule
        \end{tabular}%
    }
\end{minipage}
\vspace{-1pt}
together with their partial decryption key, $\mathsf{dk_n}$. Upon receiving all client updates, the server estimates $\mathcal{P}^r_n$ from $\mathcal{H}^r_n$ via a simple membership check across all $d$ entries of $\theta$ for all clusters. Next, clients' encrypted weights $\hat{\theta}^r_n$ are formed by replacing $\mathcal{P}_n$ indexes with the encrypted centroid values from $\hat{\mathcal{Z}}^r_n$. Finally, the server combines the received partial functional decryption keys to derive to $dk_f$ and perform a secure aggregation over clients' encrypted weights. We detail our proposed framework in Algorithm~\ref{alg:method}, while Figure~\ref{fig:overview} illustrates the~\method~training process.

\begin{algorithm}[!ht]
    \centering \scriptsize
    \caption{\small{\method: Efficient Functional Encryption in FL through weight clustering and probabilistic filters. Here, $\eta$ refers to the learning rate, while \textit{SecureAggr} refers to the secure aggregation process on (see Algorithm~\ref{alg:sec_aggr}).}}\label{alg:method}
    \begin{algorithmic}[1]
        \State \textit{DMCFE\_Init}($\lambda$, $\mathcal{N}$)
        \State Server $\mathcal{A}$ initializes model parameters $\theta$, and computes the total number of samples, $|D|$=$\sum_{i \in N} |D_i|$.
        \For{$r=1$ to $R$}
            \For{$n \in \mathcal{N}$ \textbf{in parallel}}
                \State $\left ( \mathcal{Z}^r_n, \mathcal{P}^r_n \right )$ $\gets$ \textit{ClientUpdate}($\theta^r$)
                \For{$z$ $\in$ $\mathcal{Z}^r_n$}
                    \State $\hat{z}$ $\gets$ $\mathsf{Enc}(\mathsf{ek}_n, z)$
                \EndFor
                \State $\mathcal{U}^{r}_n$ $\gets$ $\left \{ \left( i,~\mathcal{P}^{r}_{n,i} \right) \right \}_{i \in \{d\}}$
                \State $\mathcal{H}^{r}_n$ $\gets$ $\bigcup_{i \in \mathcal{U}^{r}_n} \phi(i)$ \Comment{\textit{// See Equation 2}}
                \State $\mathsf{dk}_n^r$ $\gets$ $\mathsf{dKeyShare} \left (\mathsf{sk}_n,~|D_{n}| \right )$ 
            \EndFor
            \State ${\mathsf{dk}^r}$ $\gets$ $\mathsf{dKeyComb}(\{\mathsf{dk}^r_n\}_{n \in \mathcal{N}})$
            \State $\theta^{r+1}$ $\gets$ $\textit{SecureAggr}$ $\left( \left \{ \left ( \hat{\mathcal{Z}}^{r}_n, \mathcal{H}^{r}_n \right) \right\}_{i \in \mathcal{N}},~\mathsf{dk}^r \right)$
        \EndFor
        
        \Procedure{\textit{DMCFE\_Init}}{$\lambda,~\mathcal{N}$}
            \State $\mathsf{pp} \gets \mathsf{Setup} \left(\lambda,~|\mathcal{N}|\right)$
            \For{$n \in \mathcal{N}$ clients \textbf{in parallel}}
                \State $\left ( \mathsf{ek}_n, \mathsf{sk}_n \right )$ $\gets$ $\mathsf{KeyGen}(\mathsf{id}_n)$
            \EndFor
        \EndProcedure
        \Procedure{\textit{ClientUpdate}}{$\theta$}
            \For{epoch $e=1,2,\dots,E$}
                \For{ batch $b \in \mathcal{D}_{n}$}
                    \State $\theta^{*}$ $\gets$ $\theta - \eta \cdot \nabla_{\theta} \left( \mathcal{L}_{ce} \left( \textrm{f}_{\theta}\left(b\right)\right)\right)$
                \EndFor
            \EndFor
            \State $\mathcal{Z}$ $\gets$ $\{x \mid x \in \mathrm{rand} (\theta^{*})\}_{|x|=\kappa } $ \Comment{\textit{// Cluster initialization}}
            \State ($\mathcal{Z}$, $\mathcal{P}$) $\gets$ $\mathcal{L}_{wc}(\theta^{*},~\mathcal{Z})$
        \State \Return ($\mathcal{Z}$, $\mathcal{P}$)
        \EndProcedure
    \end{algorithmic}
\end{algorithm}
\vspace{-5pt}

\subsection{Efficient Client-side Encryption}
Standard cryptographic encryption of model updates significantly increases computational overhead and communication costs, since each encrypted weight is represented by a pair of large prime numbers that are expensive to compute. Additionally, this trend exponentially escalates when security levels are increased as these prime numbers grow. To address these challenges,~\method~employs weight clustering prior to encryption, condensing model weights $\theta$ into a compact set of centroids $\mathcal{Z}_n$. This way, clients are only required to encrypt the small set of centroids, substantially lowering the computational demands and data transmission volume during the FL process. Formally, the $n$-th client aims to minimize the following loss function:

\vspace{-0.5cm}
\begin{equation}\label{eqn:client_loss}
    \begin{aligned}
        \min_{\theta} \mathcal{L}_{n}(\theta)
            &= \mathcal{L}_{wc}(\theta^{*}_{n}, \mathcal{Z}_{n}) = \sum_{j=1}^{\kappa} \sum_{i=1}^{d} u_{ij} \cdot || \theta_{n,i}^{*} - z_{n,j} ||^2~,\\
            &~\text{where}~\theta^{*}_n = \min_{\theta_n} \mathcal{L}_{ce}(\textrm{f}_{\theta_n}(\mathcal{D}_{n}))
    \end{aligned}
\end{equation}

\noindent Here, $\mathcal{L}_{ce}$ denote the cross-entropy loss computed on locally stored dataset $\mathcal{D}_n$. $Z_n$ represents the set of $\kappa$ centroids, $||\cdot||$ denotes the Euclidean distance operator, and $u_{ij}$ is a binary indicator function that returns $1$ when weight $\theta_i$ belongs to cluster $j$ and $0$ otherwise. Essentially, we first optimize model parameters $\theta$ to fit $D_n$ accurately and then we minimize the weight clustering loss, $\mathcal{L}_{wc}$, on the post-training parameters, $\theta^{*}$. 

Apart from the centroids, the cluster-weights mapping $\mathcal{P}_n$ holds crucial information about the client's model, potentially exposing sensitive data. To protect against HbC TPA and dishonest clients, we introduce an additional privacy mechanism by encoding the entries of $\mathcal{P}_n$ into a $4$-wise BF filter with 8 bits-per-parameter ($\xi=8$). Specifically, we define a set of keys $\mathcal{U}_n = \left \{ \left(i, \mathcal{P}_{n,i}\right) \mid i \in \{1, \ldots, d\} \right \}$, where each key is a pair consisting of a position $i$ within the $d$ dimensions of $\theta$ ($d$ denotes the total number of weight parameters) and the corresponding cluster identifier $\mathcal{P}_{n,i}$ in $\mathcal{P}_n$. The set of keys $\mathcal{U}_n$ are then inserted in a $4$-wise BF as per the hashing operation of Equation~\ref{eq:filter_in} to generate the fingerprint array $\mathcal{H}_n$.

The hashing operation plays a pivotal role in the transformation from $\mathcal{U}_n$ to $\mathcal{H}_n$, enhancing resistance to preimage attacks and ensuring that $\mathcal{U}_n$ cannot be deduced from $\mathcal{H}_n$. Furthermore, accurate reconstruction of the BF filter from $\mathcal{H}_n$ relies on a seed value $s$ that dictates the hashing operation's input-output relationships \textemdash~different $s$ yield varied outputs for the same input. We rely on common randomness between clients and the server (achieved through a unique seed $s_n$) to control the reconstruction of clients' BF filters, used to accurately estimate $\mathcal{P}_n$ from $\mathcal{H}_n$. While no strong security guarantees can be inferred as in cryptographic FE schemes, our mechanism effectively safeguards sensitive information against both HbC TPA and dishonest clients, avoiding the computational burden of a cryptosystem. 

\subsection{Secure Aggregation}
Once the local training at round $r$ is completed, the aggregation server $\mathcal{A}$ needs to synthesize a new global model from the information transmitted by the clients, i.e., their fingerprints, $\mathcal{H}^r_n$, and encrypted centroids, $\hat{\mathcal{Z}}^r_n$. To do so, server first reconstructs the BF filter for each client using $\mathcal{H}^r_n$ and their unique seed $s_n$. Now, the cluster-weights mappings of each client (denoted as $\mathcal{P'}^r_n$) can be estimated via a membership query for each weight across all possible clusters, as follows:

\vspace{-0.2cm}
\begin{equation}\label{eq:idx_reconstruct}
    \mathcal{P'}^r_n = \left \{ j~|~\mathrm{Member} \left (i,j \right ) = \mathrm{true} \right \}_{i \in \{d\} ,~ j \in \{\kappa\}} ~,
\vspace{0.2cm}
\end{equation}

\noindent where $\mathrm{Member}(\cdot)$ operates as defined in Equation~\ref{eq:filter_check}, with $d$ denoting the indexes of the model parameters $\theta$, and $\kappa$ referring to the cluster indexes. As the BF filters exhibit extremely small false positive rate (up to $2^{-32}$), $\mathcal{P'}^r_n$ has only a few misalignment's with the original cluster-weights mapping ($\mathcal{P'}^r_n \approx \mathcal{P}^r_n$). 

Using the estimated $\mathcal{P'}^r_n$, the server replaces the respective positions in $\mathcal{P'}^r_n$ with the corresponding encrypted centroid values from $\hat{\mathcal{Z}}^r_n$, effectively yields the encrypted client's updated model weights, denoted as $\hat{\theta}^r_n$. By combining all received clients' partial decryption keys to derive the functional decryption key, $\mathsf{dk}_f$, the server can now compute the aggregated global model as follows:

\vspace{-0.2cm}
\begin{equation}\label{eq:secure_agg}
    \theta^{r+1} = \left \{ \mathsf{Dec} \left(\{ \hat{\theta}^r_{n,i} \}_{n\in\mathcal{N}}, \mathsf{dk}_f \right ) \right \}_{i \in \{d\}}~,
\end{equation}

\noindent where $\mathsf{Dec}(\cdot)$ represents the decryption of the aggregated encrypted weights. The result forms the updated global model weights, signaling the start of the next federated round of training. Note, that in ~\texttt{EncCluster} the aggregation process can either be a plain or weighted (i.e., FedAvg) averaging of clients' model updates. In this work, we perform the latter by multiplying each clients centroids with their respective number of samples prior to encryption and scaling $\theta^{r+1}$ with the total number of samples across clients after the secure aggregation, as defined in Equation~\ref{eq:secure_agg}. An overview of the secure aggregation mechanism in~\method~can be found in Algorithm~\ref{alg:sec_aggr}.

\begin{algorithm}[!t]
    \centering \scriptsize
    \caption{\textit{SecureAggr}: Server-side secure weighted aggregation directly on encrypted client updates.}\label{alg:sec_aggr}
    \begin{algorithmic}[1]
        \State \textbf{Inputs:} Clients’ encrypted centroids and fingerprints, $\left \{ \hat{\mathcal{Z}}_n,~\mathcal{H}_n \right \}_{n \in \mathcal{N}}$, functional decryption key $\mathsf{dk}$, and total number of training samples, $|D|$.
        \State \textbf{Output:} Aggregated model parameters $\theta^{\mathrm{agg}}$.
        \For{$n \in N$}
            \State $\mathcal{P'}_n \gets \left \{ j|~\mathrm{Member} \left (i,j \right )\right \}_{ i \in \{d\},~ j \in \{\kappa\}}$ \Comment{\textit{// Equation 5}}
            \State $\hat{\theta}_n$ $\gets$ $\left \{\hat{\mathcal{Z}}_{n,i} \right \}_{ i \in \mathcal{P'}_{n}}$ \Comment{\textit{// Enc. Cluster Substitution}}
        \EndFor
        \State $\theta^{\mathrm{agg}} \gets \frac{1}{|D|}\left \{\mathsf{Dec}\left(\{\hat{\theta}_{n,i}\}_{n \in \mathcal{N}},~\mathsf{dk}\right)\right\}_{ i \in \{d\}}$
        \State \Return $\theta^{\mathrm{agg}}$
    \end{algorithmic}
\end{algorithm}

\section{Security and Privacy Analysis\label{sec:sec_analysis}}

\subsection{Security Foundations}
The security of DMCFE is crucial in~\method, forming the base of our secure aggregation. Our approach preserves DMCFE's cryptographic soundness, as shown by the formal proof in~\cite{abdalla2019decentralizing}, demonstrating ciphertext indistinguishability and adaptive corruption resistance under the DDH assumption. For more details, refer to~\cite{abdalla2019decentralizing}.

\subsection{Privacy Safeguards}
We analyze possible server active attacks, particularly isolation and replay attacks, and demonstrate how \texttt{EncCluster} effectively counters them:

\textbf{Inference Attack I (\textit{Isolation attack without collusion}):} Server $\mathcal{A}$ may attempt an isolation inference attack by selectively injecting model updates into the secure aggregation process to illicitly extract specific updates. Specifically, in round $r$, $\mathcal{A}$ seeks to decrypt and access the model updates of the $i$-th client through $\theta^{r}_i = { \mathsf{Dec}(\mathsf{dk}f, { \hat{\theta}^r{i} })}$. However, the decryption key ${dk_f}$ is generated through $\mathsf{dKeyComb}$, requiring partial decryption keys from multiple clients ($n$>$1$). This mechanism, inherent to DMCFE, ensures that isolating a single client's data for decryption is impractical, thereby effectively neutralizing the risk of this isolation attack.

\noindent \textbf{Inference Attack II (\textit{Isolation attack with collusion}):} In this scenario, server $\mathcal{A}$ conducts an isolation inference attack by collaborating with a subset $\mathcal{S} \subset \mathcal{N}$ of clients to infer the private model parameters of a target client, $n_{\mathrm{target}}$, from $\mathcal{N}$. This inference attack involves analyzing the differences between the collective raw model updates from $\mathcal{S}$ and the aggregated global model resulting from the secure aggregation mechanism. However, retrieving an individual client's local model demands the cooperation of $\mathcal{N}-1$ clients, demonstrating a significant limitation in the server's ability to conduct this attack without extensive collusion, i.e. $\mathcal{S}= \mathcal{N} \setminus \{n_{\mathrm{target}}\}$. Even when such collusion occurs in~\method, server $\mathcal{A}$ can only achieve an approximation of the client's actual weights. This limitation stems from the weight clustering process, which is tailored to each client through the objective function $\mathcal{L}_{wc}$. Consequently,~\method~ensures that accurate replication of a client's pre-clustered weights is inherently infeasible, maintaining the confidentiality of clients models. 

\textbf{Inference Attack III (\textit{Reply attack}):} A replay attack in FL involves an adversary capturing and reusing model updates from one training round in subsequent rounds, risking privacy and security by potentially exposing sensitive data or distorting the learning process. Consider three clients $i$, $j$, and $k$ engaged in two consecutive rounds, $r$ and $r+1$. Server, $\mathcal{A}$, targets participant $i$ by replaying updates from $j$ and $k$ of round $r$ in round $r+1$. Initially, secure aggregation combines updates from all three clients. However, with the replay in round $r+1$, $\mathcal{A}$ manipulates the aggregation to include previous updates from $j$ and $k$, plus the new update from $i$, allowing $\mathcal{A}$ to isolate and analyze $i$'s update changes over rounds. In~\method, client model updates are encrypted with the current round number $r$ serving as label. This effectively binds the secure aggregation process of round $r$ to the decryption keys specific to that round, thereby blocking replays and safeguarding against such privacy breaches. 

\noindent \textbf{Inference Attacks IV (\textit{Reconstruction Attack}):} Here, we explore the privacy risks inherent in weight clustering and the potential for information leakage via reconstruction attacks. Our objective is to reconstruct a client's weights, $\theta_n$, from the aggregated weights, $\theta_{\mathcal{A}}$, using supplementary information available in any aggregation round, $r$. Additionally, we note that client centroids are protected by the DMCFE cryptosystem, rendering them inaccessible to all external parties. Alternatively, attackers can extract a set of centroids by performing weight-clustering on $\theta_{\mathcal{A}}$, denoted as $\mathcal{C}_{\mathcal{A}}$, and aim to re-construct $\theta_n$ under the assumption that $\mathcal{C}_{\mathcal{A}}$ closely mirrors $\mathcal{C}_{n}$. Hence, the crux of this attack lies in finding the optimal placements of centroids $\mathcal{C}_{\mathcal{A}}$ in $\theta_n$, such that the discrepancy between estimated ($\hat{\theta}_n$) and true ($\theta_n$) client's weights distribution is minimized (e.g. $ \min \text{MSE}\left(\theta_n,~\hat{\theta}_n\right)$). However, as $\theta_n$ is unknown, $\theta_{\mathcal{A}}$ can be used as a ``\textit{best-guess}'' for the re-construction attack; thus, the goal is to minimize $\text{MSE}\left(\theta_{\mathcal{A}},~\hat{\theta}_n\right)$. In~\method, the aforementioned attack can be initiated by two distinct entities:

\vspace{-5pt}
\begin{enumerate}[label=(\alph*), wide=100pt]
    \setlength\itemsep{0em} \setlength\parsep{0em}
    \item \noindent \textbf{\textit{Malicious Clients}}: Since these parties lack the direct access to $\mathcal{P}_n$ the can only speculate on various cluster-weights mappings to approximate $\theta_n$ aiming to $\min \left(\text{MSE}\left(\theta_{\mathcal{A}},~\hat{\theta}_n\right)\right)$. However, in the absence of explicit knowledge of $\mathcal{P}_n$ the computational burden is quantified as $O(\kappa^d)$, where $d$ is the dimensionality of the client's weights, as it requires evaluating overall possible permutations between the  $\mathcal{C}_{\mathcal{A}}$ and $\theta_n$ matrices - making such process intractable as both $d$ and $\kappa$ increase.

    \item \noindent \textbf{\textit{HbC Server}}: The aggregation server $\mathcal{A}$, having access to the estimated cluster-weight mappings $\mathcal{P'}_n$, represents an internal threat model attempting to infer individual client weights $\hat{\theta}_n$ using both $C_{\mathcal{A}}$ and the estimated cluster-weight mappings $\mathcal{P'}_n$. Here, $\mathcal{A}$ has to navigate through all possible permutations between $\mathcal{C}_{\mathcal{A}}$ and $\mathcal{P'}_n$ (instead of $\theta_n$) matrices, leading to a computational complexity of $O(\kappa!)$ - a factorial increase with $\kappa$. While improved compared to $O(\kappa^d)$, such re-construction attack is still prohibited for $\kappa$>$32$, a value often used in practice. Even in the case of successful reconstruction, the server is unable to fully reconstruct client's model weights as the maximum bound of privacy leakage is given by Equation~\ref{eq:wc_privacy_aggr}, assuming $\mathcal{C}_{\mathcal{A}}$ closely mirrors $\mathcal{C}_{n}$ - an assumption often not true in pragmatic federated setting, as discussed in~\ref{asec:wc_privacy}. It is important to note, that in this case the false positive rate of the probabilistic filter encoding mechanism acts as a form of local DP mechanism, further safeguarding client-sensitive data. We perform a quantitative evaluation in Section~\ref{sec:quant_wc}.
\end{enumerate}

\vspace{-10pt}
\noindent \textbf{Inference Attack V (\textit{Compromised TPA}):} In~\method, the role of a fully trusted authority for initializing security ($\lambda$) and public parameters ($pp$), along with coordinating client-specific key generation ($sk$,$ek$) via identifiers ($\mathsf{id}$), is re-evaluated. Considering the TPA as a potentially HbC entity, it can use $\mathsf{id}$'s to access clients' $sk$ and $ek$. If the TPA intercepts communications from a specific client $j$, it could decrypt $Z^{r}_j$ by crafting ``\textit{dummy}'' updates for all other clients and conducting a secure aggregation. Nevertheless, access to $j$'s raw centroid values does not enable the TPA to accurately reconstruct $\theta^{r}_j$. The reconstruction is hindered by the complexity of deriving the cluster-to-weights mapping $\mathcal{U}_n$ from $\mathcal{H}^{r}_j$, compounded by BF filters hashing operations' sensitivity to initial unknown seeding ($s_j$), preserving data integrity against a compromised TPA. Moreover, the TPA can attempt to estimate a client's model parameters using $Z^{r}_j$ and the round's aggregated model $\theta^r$ by replacing each weight in $\theta^r$ with its nearest centroid in $Z^{r}_j$. However, the variability in weight clustering optimization (Eq.~\ref{eqn:cluster}) among clients, as seen in weight clustering FL approaches~\cite{khalilian2023fedcode}, results in flawed estimations, highlighting~\method's defense against a compromised TPA.

\subsubsection{{Quantitative Evaluation on Privacy Leakage due to Weights Clustering}\label{sec:quant_wc}}

We now evaluate the ability of server to re-construct individual clients weights by performing an cluster inference attack as presented in Inference Attacks IV. For this, we performed experiments, where we measured the similarity between the ``\textit{perfectly}'' estimated client weights (e.g. replacing each entry in $\theta_n$ with their closest value in $\mathcal{C}_{\mathcal{A}}$), $\hat{\theta}_n$, and the true client's weights, $\theta_n$, by measuring the similarity of the two models in the embeddings space. Specifically, we extract embeddings on the client's locally stored data using both $\hat{\theta}_n$ and $\theta_n$, after which we perform a dimensionality reduction through PCA and measure the MSE error, similar to~\cite{wei2020framework}. 

\begin{figure}[!t]
  \centering
  \begin{subfigure}[b]{0.45\textwidth}
    \includegraphics[width=\textwidth]{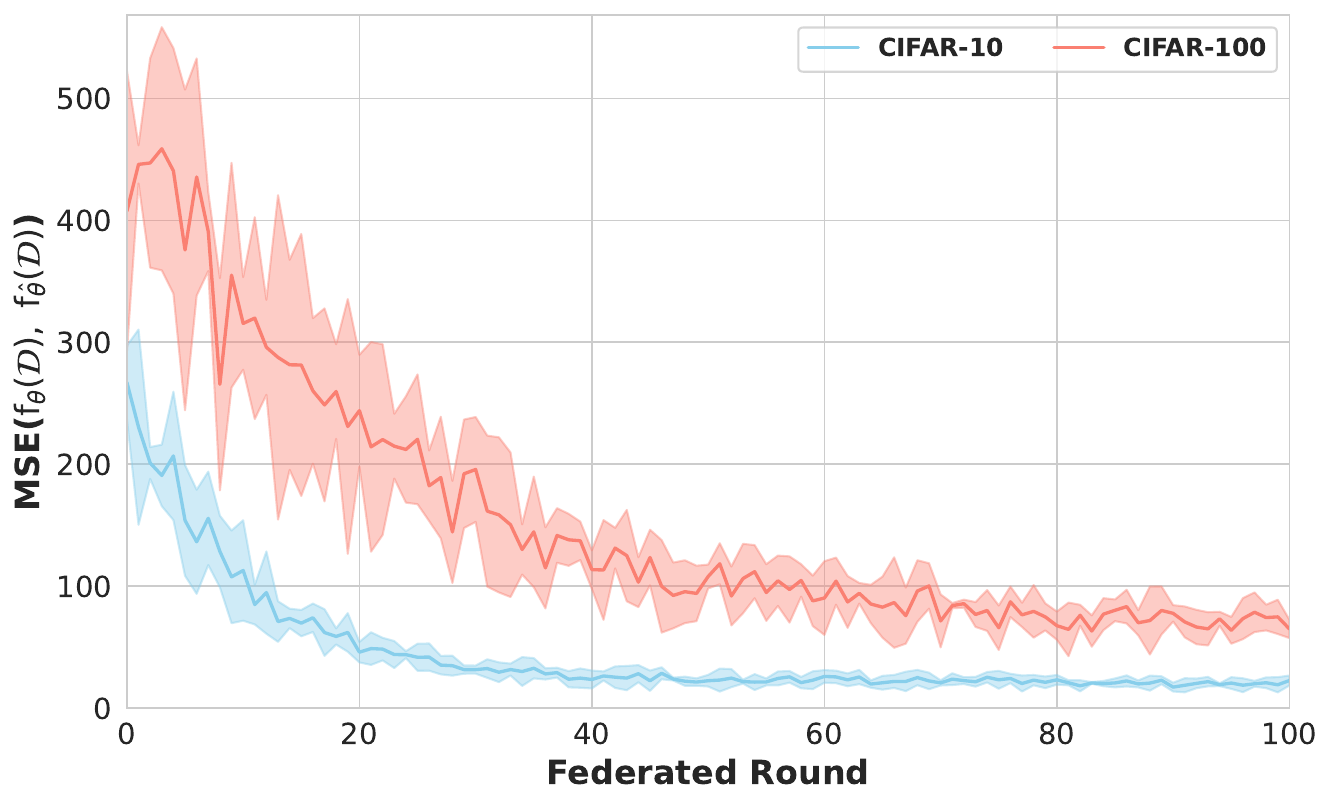}
    \caption{IID settings} \label{fig:mse_iid}
  \end{subfigure}
  \hfill
  \begin{subfigure}[b]{0.45\textwidth}
    \includegraphics[width=\textwidth]{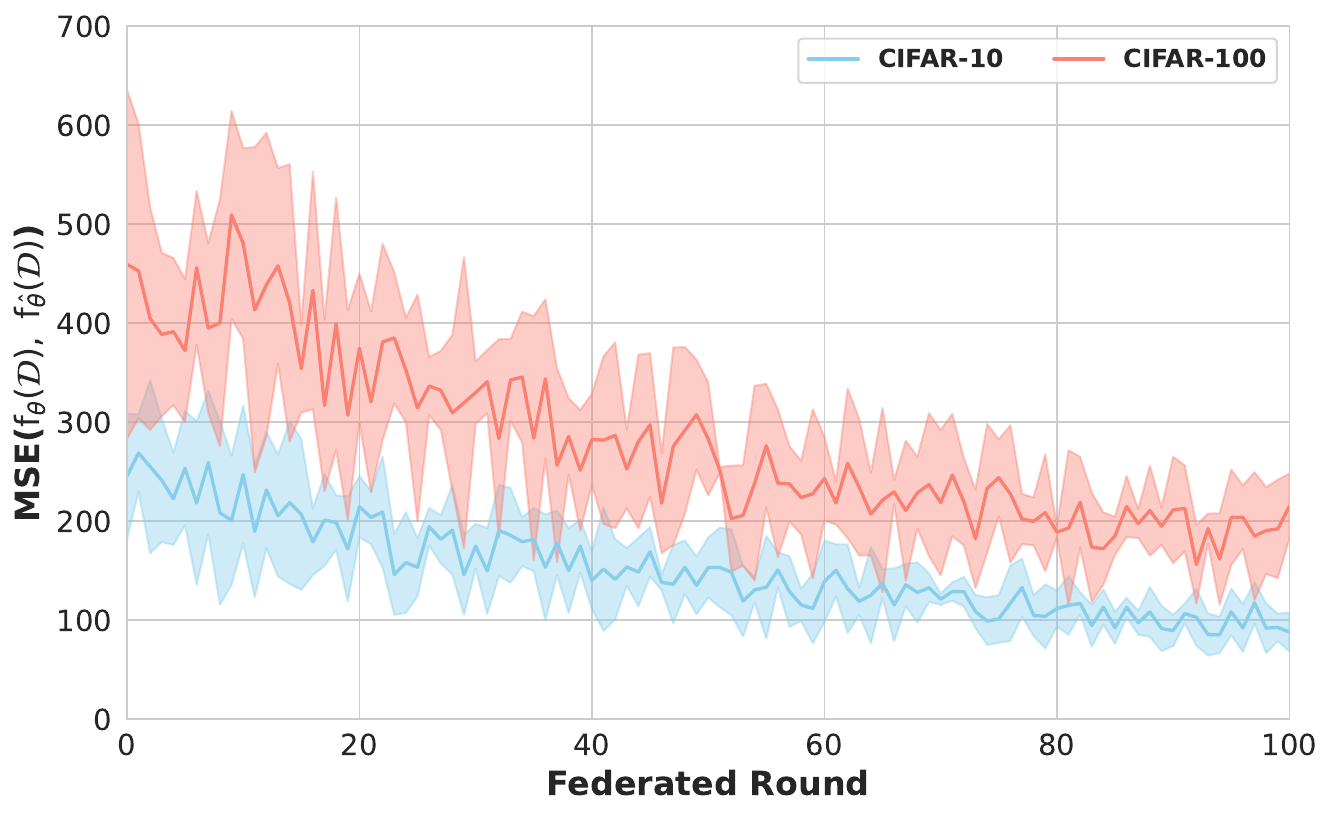}
    \caption{Non-IID settings} \label{fig:mse_noniid}
  \end{subfigure}
  \caption{Evaluation of Cluster Inference Attacks in~\method. We report MSE
  between the client's data embeddings extracted from client's true and estimated weights for both (a) IID and (b) non-IID settings.} \label{fig:emb_mse}
\end{figure}

In Figure~\ref{fig:emb_mse}, a pronounced contrast emerges between IID and non-IID settings. Specifically, in IID scenarios where there's a significant overlap between $\mathcal{C}_{\mathcal{A}}$ and $\mathcal{C}_{n}$, the attacker achieves notably lower MSE values, particularly in the later stages of training. Conversely, in non-IID settings, the effectiveness of the attack diminishes significantly, marked by high fluctuations across training rounds. Given the prevalence of highly non-IID conditions in most practical FL environments, Figure~\ref{fig:mse_noniid} underscores the potential for privacy breaches stemming from weight clustering in \method. Nonetheless, it's critical to acknowledge that these findings are predicated on the assumption of a ``\textit{perfect}'' estimation scenario, where $\theta_n$ entries are precisely matched with the nearest values in $\mathcal{C}_{\mathcal{A}}$, necessitating $O(\kappa^d)$ complexity. This scenario underscores that, even when subjected to such potent attack strategies, the server's capacity to accurately reconstruct client models is limited, thereby offering a degree of protection against private data exposure.

\section{Efficiency Evaluation\label{sec:exp}}


\subsection{Experimental Setup}

\noindent \textbf{Datasets \& Models.} We use publicly available datasets from vision domain, namely CIFAR-10 and CIFAR-100 \cite{krizhevsky2009learning}. In our experiments, we used AutoAugment \cite{cubuk2018autoaugment} for our data augmentation and evaluated~\method~across 4 distinct neural network architectures, namely ResNet-20~\cite{resnet}, ConvMixer-256/8~\cite{trockman2022patches}, ConvNeXt-Tiny~\cite{woo2023convnext} and MobileNet~\cite{sandler2019mobilenetv2}, training all models from scratch (no pre-trained weights).

\noindent \textbf{Federated Settings.} We used Flower~\cite{beutel2020flower} to simulate a federated environment, controlling the setup with the following parameters: number of clients - $M$, rounds - $R$, local train epochs - $E$, client participation rate per round - $\rho$, client class concentration - $\gamma$, and number of clusters - $\kappa$. For our experiments, we set $M$=$30$, $E$=$1$, and $\kappa$ to $128$ (except for experiments in Figure~\ref{fig:clusters_all}). For $\rho=1$, we used $R=100$, while for $\rho<<1$, $R$ was set to $300$ for IID and $500$ for non-IID experiments with random client selection each round. Unless specified otherwise, a key size ($\textrm{KS}$) of $256$ was used for DMCFE.

\noindent \textbf{Baselines.} We evaluate~\method~in terms of final model accuracy, computational complexity of the encryption process, and data transmitted compared to standard \textit{FedAvg} (denoted as \textit{$\times$times FedAvg}). We include FedAvg$_{wc}$ (\textit{FedAvg} with client-side weight clustering) to highlight the effect of weight clustering on model performance. We compare with SEFL~\cite{mohammadi2023secure}, which combines Paillier Homomorphic Encryption (PHE) with gradient magnitude pruning to reduce computational burdens on clients — a strategy similar to ours. In SEFL, we prune 70\% and 60\% of gradients for CIFAR-10/100 respectively (found by initial exploration), while we use Huffman encoding on sparse pruned weights to further minimize communication overhead. Additionally, we include DeTrust-FL~\cite{xu2022detrust} to illustrate the benefits of our framework against other FE adaptations in FL.

\subsection{FL training on Data Splits}~\label{ssec:splits} \vspace{-0.5cm}

\noindent \textbf{IID Data Split.} Here, we focus on IID data distribution, with the number of clients ($N$) set to 30 and under full ($\rho$=$1$) and partial ($\rho$=$0.2$ \textemdash~meaning each round $\rho \cdot N = 6$ clients are randomly selected) client participation rate. As depicted in Table~\ref{tab:main_res},~\method~effectively minimizes accuracy loss while significantly enhancing communication efficiency and reducing computational complexity on the client side. Compared to DeTrust-FL \textemdash~which employs the same FE cryptosystem \textemdash~\method~dramatically reduces the communication burden (up to approximately $13\times$ less) and slashes encryption time from over $325$ seconds to just $2$ seconds, consistently across all experiments. Against SEFL, which utilizes gradient pruning and HE for improved efficiency,~\method~exhibits superior model performance with an average accuracy loss of -0.91\% compared to SEFL's -1.72\%. This superiority is further highlighted in comparison with FedAvg$_{wc}$, which, like~\method, outperforms SEFL in terms of accuracy loss. Furthermore, \method~achieves substantial reductions in both communication costs and encryption time relative to SEFL, with data volume decreasing from $2.6$ to $0.28$ times that required by \textit{FedAvg}, and encryption time reduced by approximately a factor of $4$. Crucially, \method~maintains its superior performance even under lower participation rates ($\rho=0.2$), ensuring strong privacy guarantees without the high computational and communication burdens typically associated with FE, all while minimally impacting model performance. \\

\noindent \textbf{Non-IID Data Split.} We now evaluate in a more realistic federated setting, where clients' data follow a non-IID distribution using $Dir(0.1)$ over classes ($\gamma \approx 0.2$). Here, we maintain the remaining parameters identical to the IID case ($N=30$, $\rho \in [0.2,1.0]$), enabling us to analyze~\method~effectiveness in complex and realistic conditions where clients have limited network resources and exhibit significantly varied data distributions. From the non-IID column of Table~\ref{tab:main_res}, we notice similar gains over the baselines as the IID experiments; in that~\method~can achieve substantial reductions in both computational and communication demands while upholding strong privacy guarantees akin to those offered by DeTrust-FL, without compromising on model performance. Specifically, while SEFL encounters difficulties in maintaining optimal model performance, suffering an accuracy loss of up to approximately 4\%,~\method~maintains a $\delta$-Acc under 2\% while simultaneously achieves substantial efficiency gains, as evidenced by 9-fold and 4-fold reductions in encryption time and communication costs, respectively. This behavior becomes even more pronounced under conditions of partial client participation ($\rho=0.2$) in more complex tasks, such as CIFAR-100, where the disparity in model performance between the approaches extends to 2\%.

\begin{table}[!t]
    \centering
    \caption{Performance evaluation of~\method~in both IID \textemdash~$\gamma \approx 1.0$ using $Dir$($10$) over classes \textemdash~and non-IID \textemdash~$\gamma \approx 0.2$ using $Dir$($0.1$) over classes \textemdash~settings across different datasets using ResNet-20. We report the accuracy loss ($\delta$-Acc) on test set, data transmitted in upstream communication route compared with \textit{FedAvg}, and client-side encryption times. For ease of comparison, we include standard \textit{FedAvg} accuracy on test set, denoted as Accuracy.} \label{tab:main_res}
    \resizebox{0.8\textwidth}{!}{%
        \begin{tabular}{cclcccccc}
            \toprule

            & \multicolumn{1}{c}{\multirow{2}{*}{\textbf{Dataset}}}
            & \multicolumn{1}{c}{\multirow{2}{*}{\textbf{Approach}}}
            & \multicolumn{2}{c}{\textbf{IID ($\gamma \approx 1.0$)}}
            & \multicolumn{2}{c}{\textbf{non-IID ($\gamma \approx 0.2$)}}
            & \multicolumn{1}{c}{\multirow{2}{*}{\begin{tabular}[c]{@{}c@{}}\textbf{\textit{Data Transmitted}}\\\textbf{\textit{($\times$~times FedAvg)}}\end{tabular}}}
            & \multicolumn{1}{c}{\multirow{2}{*}{\begin{tabular}[c]{@{}c@{}}\textbf{\textit{Client-side}}\\\textbf{\textit{Encryption (s)}}\end{tabular}}}
            \\
            \cmidrule(l{.01\linewidth} r{.001\linewidth}){4-5}
            \cmidrule(l{.01\linewidth} r{.001\linewidth}){6-7}
            & & 
                & \textbf{\textit{Accuracy}} & \textbf{\textit{$\delta$-Acc.}}
                & \textbf{\textit{Accuracy}} & \textbf{\textit{$\delta$-Acc.}}
            & & \\
            \midrule

            \multicolumn{1}{c}{\multirow{10}{*}{$\mathbf{\rho}=1$}}
            & \multirow{5}{*}{\textbf{CIFAR-10}}
                & \textbf{FedAvg$_{wc}$} & \multirow{5}{*}{\textit{89.07}}
                                                            & -0.12 & \multirow{5}{*}{\textit{83.12}} & -0.71 & 0.034 & \textemdash \\
                & & \textbf{DeTrust-FL} &                   & -0.09 &                                 & -0.10 & 3.743 & 329.53 \\ 
                & & \textbf{SEFL} &                         & -0.98 &                                 & -1.74 & 2.581 & 7.64 \\
                & & \textbf{EncCluster (Ours)} &            & \textbf{-0.32} &                        & \textbf{-0.79} & \textbf{0.284} & \textbf{2.04} \\
            \cmidrule{2-9}

            & \multirow{5}{*}{\textbf{CIFAR-100}}
                & \textbf{FedAvg$_{wc}$} & \multirow{5}{*}{\textit{61.33}}
                                                            & -1.07 & \multirow{5}{*}{\textit{54.37}}   & -1.64 & 0.035 & \textemdash \\
                & & \textbf{DeTrust-FL} &                   & -0.11 &                                   & -0.09 & 3.757 & 338.61 \\ 
                & & \textbf{SEFL} &                         & -2.42 &                                   & -3.88 & 2.656 & 8.01 \\
                & & \textbf{EncCluster (Ours)} &            & \textbf{-1.21} &                          & \textbf{-1.67} & \textbf{0.285} & \textbf{2.03} \\
            \midrule

            \multicolumn{1}{c}{\multirow{10}{*}{$\mathbf{\rho}=0.2$}}
            & \multirow{5}{*}{\textbf{CIFAR-10}}
                & \textbf{FedAvg$_{wc}$} & \multirow{5}{*}{\textit{88.42}}
                                                            & -0.28 & \multirow{5}{*}{\textit{81.91}}   & -0.89 & 0.032 & \textemdash \\
                & & \textbf{DeTrust-FL} &                   & -0.06 &                                   & -0.12 & 3.744 & 326.15 \\
                & & \textbf{SEFL}   &                       & -1.14 &                                   & -1.97 & 2.569 & 7.63 \\
                & & \textbf{EncCluster (Ours)} &            & \textbf{-0.44} &                          & \textbf{-1.05} & \textbf{0.281} & \textbf{2.04} \\
            \cmidrule{2-9}

            & \multirow{5}{*}{\textbf{CIFAR-100}}
                & \textbf{FedAvg$_{wc}$} & \multirow{5}{*}{\textit{60.07}}
                                                        & -1.32 & \multirow{5}{*}{\textit{47.37}}   & -1.87 & 0.033 & \textemdash \\
                & & \textbf{DeTrust-FL} &               & -0.09 &                                   & -0.03 & 3.756 & 339.12 \\
                & & \textbf{SEFL} &                     & -2.35 &                                   & -3.79 & 2.632 & 7.98 \\
                & & \textbf{EncCluster (Ours)} &        & \textbf{-1.67} &                          & \textbf{-2.03} & \textbf{0.279} & \textbf{2.03} \\
            \bottomrule

        \end{tabular}%
    }
\end{table}

\subsection{\method~Component Analysis}~\label{ssec:blocks} \vspace{-20pt}

Now, we perform an ablation study on fundamental components of~\method: the encoding of cluster-weights mappings using BF filters, and the cluster size $\kappa$ employed in the weight clustering process, with the aim to analyze how these elements affect the model's performance and efficiency.

\noindent \textbf{BF filter Efficiency.} To assess the impact of Binary Fuse (BF) filters on computational complexity, communication overhead, and model accuracy loss ($\delta$-Acc), we conducted experiments on ResNet-20 with $N$=$30$, comparing a variant of our method, denoted as~\method$_{noBF}$, which forgoes BF hashing operations in favor of Huffman encoding for transmitting cluster-weights mappings. Table~\ref{tab:bf_effect} demonstrates that BF filters minimally affect model performance, a consistency observed across our experiments. This is attributed to the near-perfect reconstruction of mappings on the server side, facilitated by the BF filters' extremely low false positive rates. Moreover, comparing encryption times between~\method$_{noBF}$ and~\method~shows that BF filters add minimal computational overhead, with encryption time slightly increasing from 1.72 to approximately 2 seconds. In terms of communication overhead relative to \textit{FedAvg},~\method$_{noBF}$ requires only 0.032 times the data volume, whereas incorporating BF filters increases this to 0.284, due to the bit requirement per filter entry (approximately $8.68$ bits-per-entry). Despite this increase,~\method~still achieves significant reductions in communication costs compared to other privacy-preserving adaptations in FL, as detailed in Table~\ref{tab:main_res}, all while enhancing privacy guarantees within such systems. Significantly,~\texttt{EncClu} \texttt{ster}$_{noBF}$ proves particularly advantageous in FL systems with limited communication resources operating under less ``\textit{stringent}'' threat models \textemdash~such as when a fully-trusted authority exists \textemdash~delivering computational efficiency while substantially reducing communication overhead to over $100$ $\times$ less than DeTrust-FL's. 

\begin{table}[!t]
    \centering
    \caption{Effect of BF filters in~\method. Experiments conducted using ResNet-20 with $N=30$. We report the accuracy loss ($\delta$-Acc) on test set, data transmitted in upstream communication route compared with \textit{FedAvg}, and client-side encryption times.}\label{tab:bf_effect}
    \resizebox{0.7\textwidth}{!}{%
        \begin{tabular}{lccccc}
            \toprule
            & \multirow{2}{*}{$\mathbf{\rho}$}
            & \multicolumn{2}{c}{\textbf{CIFAR-10}} & \multicolumn{2}{c}{\textbf{CIFAR-100}} \\
            \cmidrule(lr){3-4} \cmidrule(lr){5-6}
                            & & \textbf{\textit{EncCluster}$_{noBF}$
                            } & \textbf{\textit{EncCluster}} & \textbf{\textit{EncCluster}$_{noBF}$} & \textbf{\textit{EncCluster}} \\
            \midrule
            \multirow{2}{*}{\begin{tabular}[c]{@{}c@{}}\textbf{\textit{IID}}\\($\gamma \approx 1.0$)\end{tabular}} 
                                        & $0.2$ & -0.31 & -0.44 & -1.31 & -1.67 \\
                                        & $1$   & -0.14 & -0.32 & -1.09 & -1.21 \\
            \midrule
            \multirow{2}{*}{\begin{tabular}[c]{@{}c@{}}\textbf{\textit{non-IID}}\\($\gamma \approx 0.2$)\end{tabular}}
                                        & $0.2$ & -0.94 & -1.05 & -1.85 & -2.03 \\
                                        & $1$   & -0.73 & -0.79 & -1.63 & -1.67\\
            \midrule
            \multicolumn{2}{l}{\multirow{2}{*}{\textbf{\textit{Encryption time (s)}}}} & \multirow{2}{*}{1.72} & \multirow{2}{*}{2.04} & \multirow{2}{*}{1.72} & \multirow{2}{*}{2.03} \\ \\
            \multicolumn{2}{c}{\textbf{\begin{tabular}[c]{@{}c@{}}\textbf{\textit{Data Transmitted}}\\(\textit{$\times$~FedAvg})\end{tabular}}}
            & 0.034 & 0.284 & 0.035 & 0.281 \\
            \bottomrule
        \end{tabular}%
    }
\end{table}

\begin{figure*}[!b]
    \centering
    \begin{subfigure}[b]{0.45\textwidth}
       \centering \includegraphics[width=0.99\textwidth]{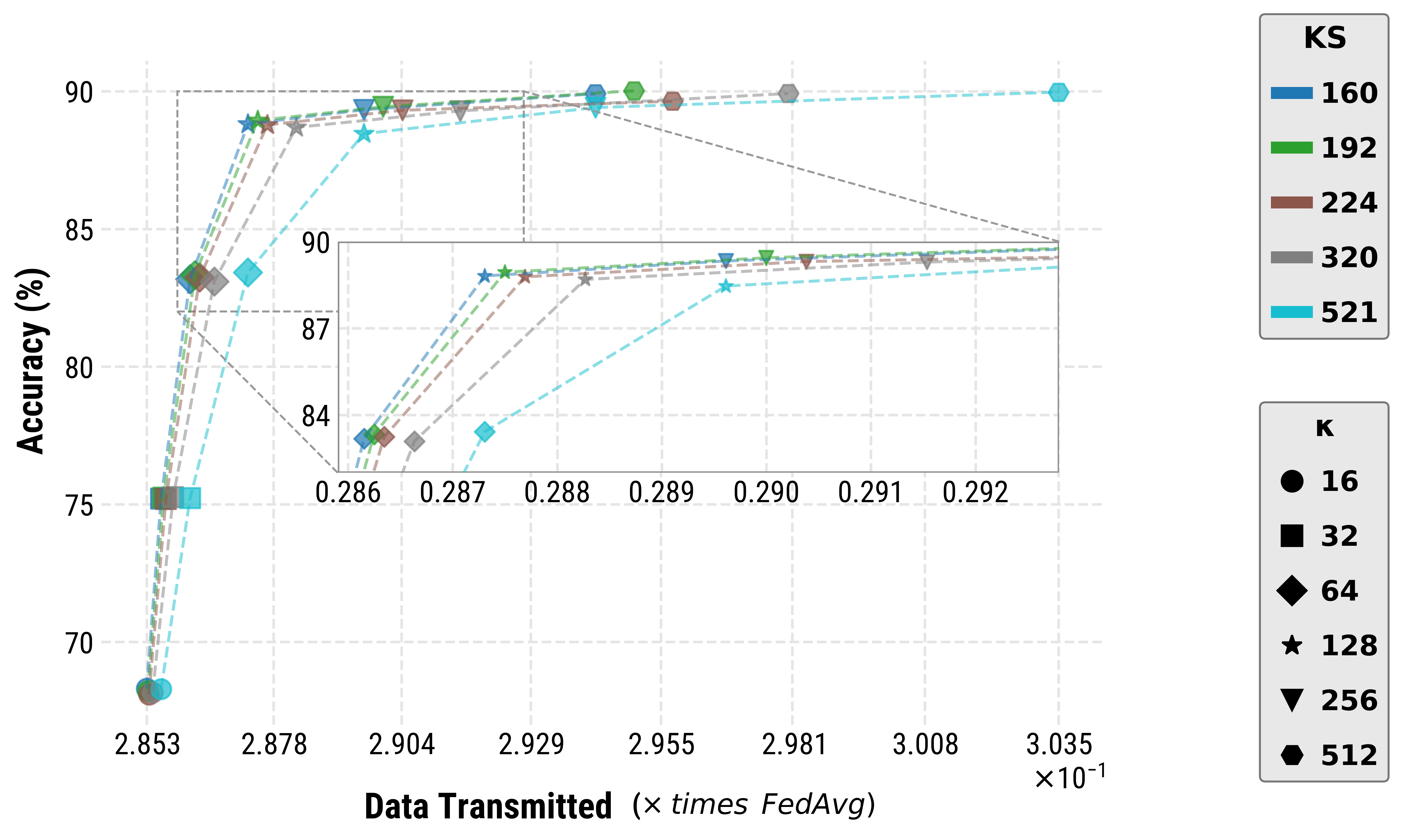}
        \caption{Impact of cluster size ($\kappa$) and DMCFE key ($\textrm{KS}$).
        }\label{fig:clusters}
    \end{subfigure}
    \hfill
    \begin{subfigure}[b]{0.45\textwidth}
        \centering \includegraphics[width=0.99\textwidth]{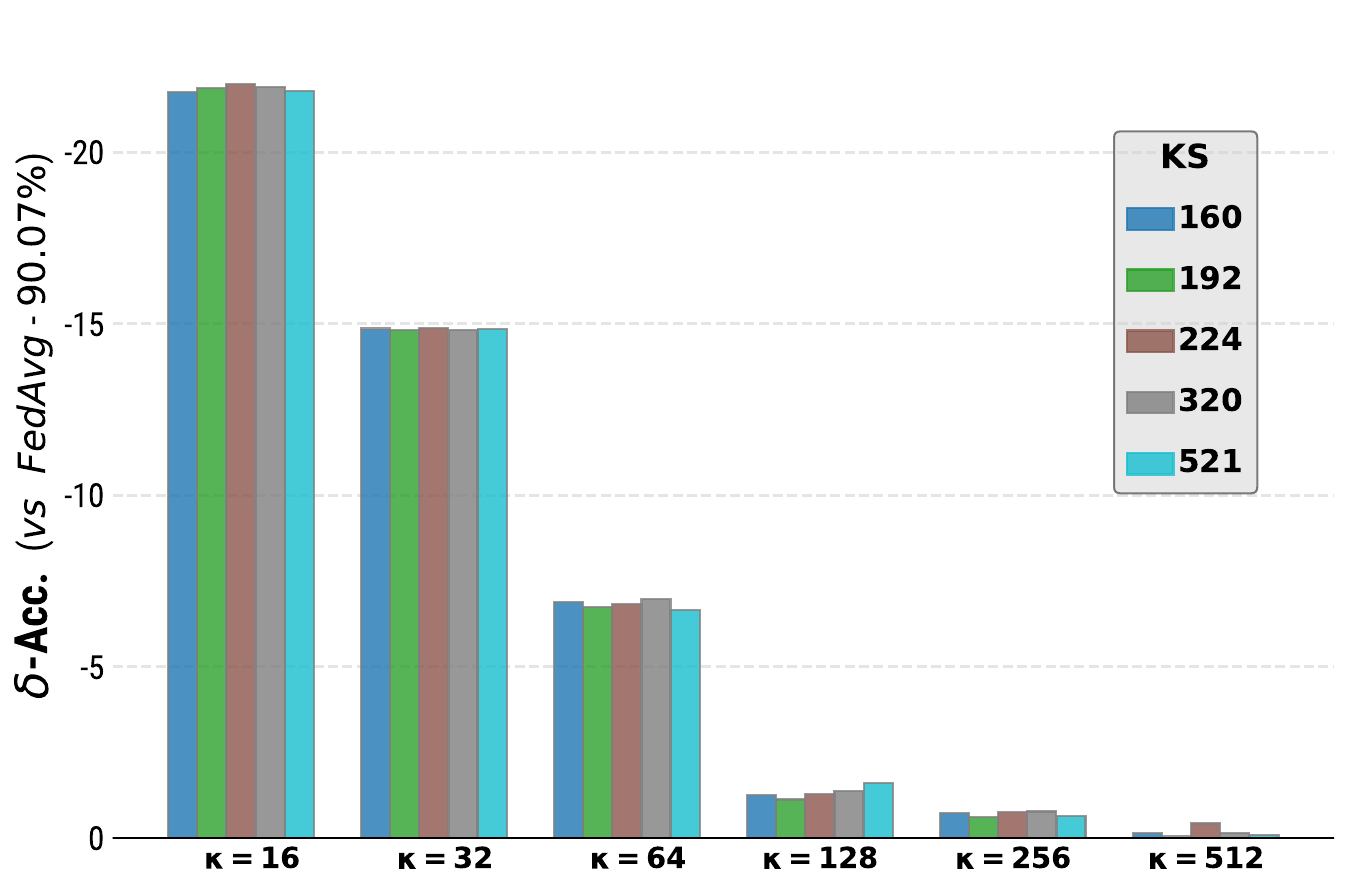}
        \caption{Accuracy loss ($\delta$-Acc.) across cluster sizes ($\kappa$).}\label{fig:clusters_zoomed}
    \end{subfigure}
    \caption{\method~performance evaluation with varied cluster sizes ($\kappa$) and DMCFE key sizes ($\textrm{KS}$). Experiments conducted using ResNet-20 under IID settings ($\gamma \approx 1.0$) on the CIFAR-10. Subfigure (a) shows model accuracy on the test set and data transmitted in upstream communication compared with \textit{FedAvg}, while (b) illustrates accuracy loss versus \textit{FedAvg} for various $\kappa$ and $\textrm{KS}$ values. Federated parameters were set to $N=10$, $R=100$, and $\rho=1.0$.}\label{fig:clusters_all}
\end{figure*}

\noindent \textbf{Cluster Size Effect.} To investigate the impact of cluster size, $\kappa$, on model performance in~\method, we conducted experiments across a range of $\kappa$ values from 16 to 512, with $N=10$ and full client participation in IID settings ($\gamma \approx 1.0$) on CIFAR-10. Concurrently, we varied the key size ($\textrm{KS}$) within the DMCFE cryptosystem to explore the relationship between cluster size and communication overhead under different security levels, as indicated by changes in $\textrm{KS}$. Figure~\ref{fig:clusters} reveals that model accuracy is significantly affected by cluster size, rising from approximately 68\% at $\kappa=16$ to over 89\% at $\kappa=512$. This trend highlights the role of weight clustering's cluster size, $\kappa$, in effectively capturing the nuances of clients' post-training model parameters, as explained in Equation~\ref{eqn:cluster}. Notably, this improvement in accuracy is observed consistently across various $\textrm{KS}$ values \textemdash~as evident in Figure 2b \textemdash~suggesting that model performance is primarily influenced by $\kappa$, rather than the level of encryption. However, a plateau in accuracy improvements is observed at $\kappa=128$; beyond this point, further increases cease to provide proportional benefits. Instead, higher communication costs arise, alongside additional computational overhead due to the computational complexity of clustering, denoted as $\mathcal{O}(\kappa \cdot d)$. Interestingly, the level of security minimally impacts data volume, with only slight variations between 0.285 and 0.303 across all $\textrm{KS}$ values, owing to the encryption of solely $\kappa$ values \textemdash~a minor portion of the total transmitted data. Consequently,~\method's clustering approach allows for efficient scaling to higher encryption levels without significantly escalating communication overhead during FL training, presenting a scalable FE scheme for FL.

\subsection{Computational Speed-up and Communication Reduction}~\label{ssec:enc_level} \vspace{-25pt}

\begin{figure*}[!t]
    \centering
    \begin{subfigure}[b]{0.96\textwidth}
        \includegraphics[width=\textwidth]{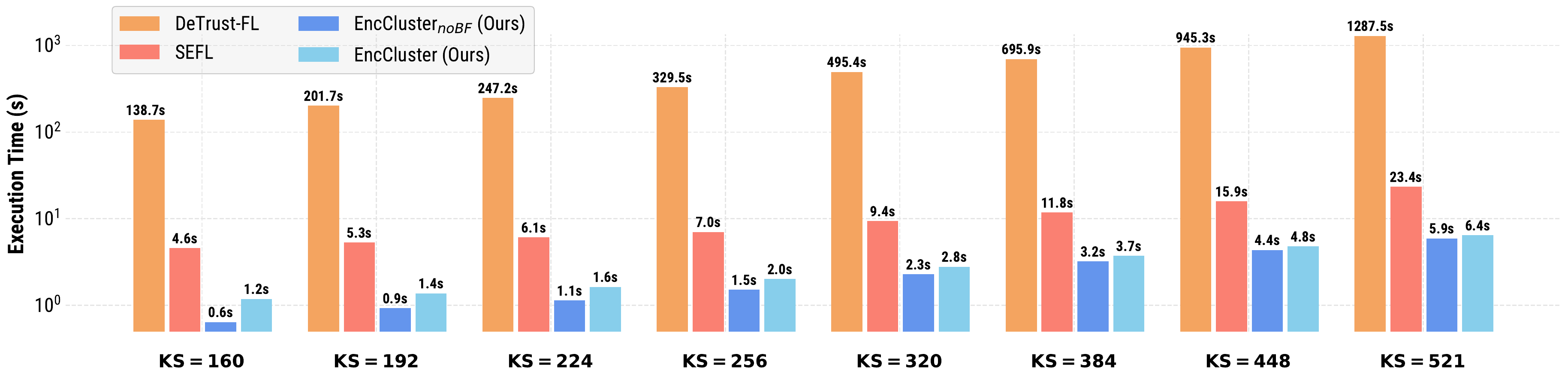}
        \caption{Execution time of encryption process.}\label{fig:exec_time}
    \end{subfigure}
    \hfill
    \begin{subfigure}[b]{0.96\textwidth}
        \includegraphics[width=\textwidth]{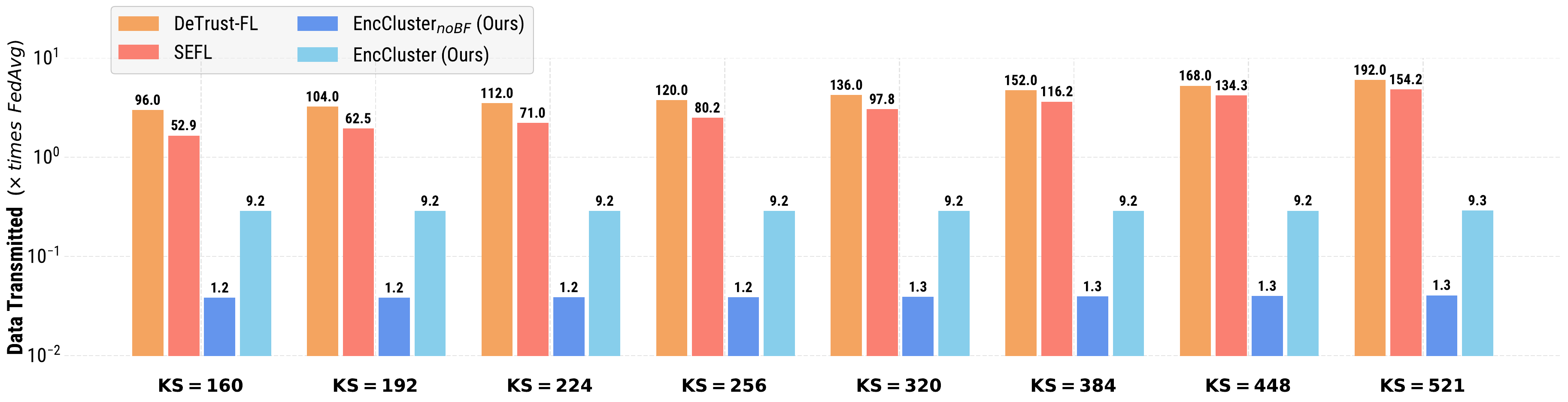}
        \caption{Volume of transmitted data compared to standard \textit{FedAvg}.} \label{fig:communcation}
        \vspace{-8pt}
    \end{subfigure}
    \caption{Evaluation of computation and communication efficiency for~\method. Experiments conducted with ResNet-20 under IID settings ($\gamma \approx 1.0$) on CIFAR-10 with $N=10$ and $\rho=1$. In (a), we detail the computational time required for the encryption process in a logarithmic scale, while (b) contrasts the volume of data transmitted with that of the standard \textit{FedAvg}. In (b), we specify the average bits-per-parameter ($bpp$), indicated by the numerals atop each experiment to clearly illustrate~~\method~efficiency.\label{fig:rel_data}}
    \vspace{-12pt}
\end{figure*}

To investigate the scalability of~\method~across varying encryption levels, we perform experiments with various key sizes ($\textrm{KS}$) within the DMCFE cryptosystem using ResNet-20 where $N=10$ and $\rho=1$ in IID settings ($\gamma \approx 1.0$) on CIFAR-10. With the increase in $\textrm{KS}$ leading to more complex computations and larger encrypted messages, potentially impacting efficiency, we report both the encryption time and communication cost during training (quantified as $\times$ \textit{FedAvg}'s data volume). To clearly demonstrate the effectiveness of~\method, we conduct identical experiments on all our baselines and report our findings in Figure~\ref{fig:rel_data}. 

\noindent \textbf{Computational Speedup.} Figure~\ref{fig:exec_time} reveals that~\method~consistently maintains low execution times across all key sizes, with a maximum of only $6.44$ seconds at $KS=521$. In contrast, DeTrust-FL shows a significant increase in encryption time with larger $\textrm{KS}$ values, highlighting its scalability issues. SEFL, while more efficient than DeTrust-FL, remain approximately 4 $\times$ slower than~\method~across all experiments. Analyzing both~\method~variants, namely~\method$_{noBF}$ and~\method, we observe a marginal overhead introduced by BF filters fixed across all key sizes, suggesting that BF filters' privacy benefits come with negligible computational overhead. These findings underscore~\method~computational efficiency, marking it suitable for secure FL even with increased encryption demands. 

\noindent \textbf{Communication Reduction.} In Figure~\ref{fig:communcation}, we notice that~\method~maintains a near-constant communication cost \textemdash~substantially smaller than both DeTrust-FL and SEFL \textemdash~across all experiments. This is primarily attributed to the weight clustering procedure that results in encryption of solely $\kappa$ values, adding negligible overhead to the total data transmitted. Notably,~\method~offers superior communication efficiency even when compared to the standard \textit{FedAvg}, which lacks privacy guarantees against inference attacks, thereby enhancing security in FL while reducing its communication burdens. Furthermore,~\method$_{noBF}$ showcases a dramatic reduction in communication volume, exponentially more efficient to all baselines, underscoring its applicability for FL systems under strict bandwidth constraints that require robust privacy protections.

\begin{table*}[!t]
    \centering
    \caption{\method~performance across various neural network architectures under IID settings ($\gamma \approx 1.0$) on the CIFAR-10 dataset. We report the standard performance of \textit{FedAvg} (\textit{Accuracy}) and the accuracy loss ($\delta$-Acc) introduced by~\method, as well as the client-side encryption time (measured in seconds). Federated parameters were set to $N=10$, $R=100$, and $\rho=1.0$.}\label{tab:model_var}
    \vspace{0.1cm}
    \resizebox{0.82\textwidth}{!}{%
        \begin{tabular}{lcccccccc}
            
            \toprule
            \multicolumn{1}{c}{\multirow{2}{*}{\textbf{Dataset}}} &
            \multicolumn{2}{c}{\multirow{1}{*}{\textbf{ConvNeXt-Tiny}}} &
            \multicolumn{2}{c}{\multirow{1}{*}{\textbf{ResNet-20}}} &
            \multicolumn{2}{c}{\multirow{1}{*}{\textbf{MobileNet}}} &
            \multicolumn{2}{c}{\multirow{1}{*}{\textbf{ConvMixer-256/8}}}
            \\
            \cmidrule(l{.01\linewidth} r{.001\linewidth}){2-3} 
            \cmidrule(l{.01\linewidth} r{.001\linewidth}){4-5} 
            \cmidrule(l{.01\linewidth} r{.001\linewidth}){6-7}
            \cmidrule(l{.01\linewidth} r{.001\linewidth}){8-9}
            
            & \textbf{\textit{Accuracy}} & \textbf{\textit{$\delta$-Acc.}} &
            \textbf{\textit{Accuracy}} & \textbf{\textit{$\delta$-Acc.}} &
            \textbf{\textit{Accuracy}} & \textbf{\textit{$\delta$-Acc.}} &
            \textbf{\textit{Accuracy}} & \textbf{\textit{$\delta$-Acc.}} 
            \\ \midrule

            \textbf{CIFAR-10} 
                & 86.47 & -0.36 
                & 89.07 & -0.32
                & 91.48 & -0.24 
                & 92.35 & -0.19 
                \\

            \textbf{CIFAR-100}  
                & 60.35 & -1.29 
                & 61.33 & -1.19
                & 70.02 & -0.97
                & 72.64 & -1.02 
                \\ \midrule
            \multicolumn{1}{l}{\multirow{1}{*}{\textbf{\textit{Encryption time (s)}}}} & 
                \multicolumn{2}{c}{2.02} & 
                \multicolumn{2}{c}{2.03} &
                \multicolumn{2}{c}{2.08} &
                \multicolumn{2}{c}{2.12} 
                \\
            \bottomrule
        \end{tabular}%
    }
\end{table*}

\subsection{Generalization Across Neural Architectures}~\label{ssec:var_arch} \vspace{-0.5cm}

Here, we assess~\method's performance across various neural architectures, examining accuracy loss and encryption time. We exclude communication reduction results, as their outcomes, tied to the weight clustering process, mirror findings from previous discussions. Instead, we report encryption time, as it accounts for the injection of cluster-weights mappings into BF filters \textemdash~a process influenced by the model’s parameter count, ranging from $0.16$M in ConvNeXt-Tiny to $3.4$M in MobileNet. In Table~\ref{tab:model_var},~\method~demonstrates consistently minimal accuracy loss across all architectures and datasets. Remarkably, for larger models like MobileNet and ConvMixer-256/6,~\method~exhibits even less impact on performance, with $\delta$-Acc as low as -0.19 for CIFAR-10 and -0.97 for CIFAR-100. Furthermore, encryption times confirm~\method's high computational efficiency irrespective of the model complexity, as the number of parameters to be encrypted remains constant ($\kappa$), and BF filter hashing operations are low-cost, with only a 0.1-second increase for key injections ranging from $0.16$M to $3.4$M. These findings demonstrate \method's seamless integration across different architectures, ensuring minimal performance loss while providing significant computational and communication efficiencies, all essential for scalable and secure FL. 

\vspace{-10pt} \section{Related Work\label{sec:related}}

\noindent \textbf{Privacy-preserving FL.} Vanilla FL schemes are prone to privacy concerns, as they are susceptible to inference attacks during model updates between clients and the aggregation server~\cite{nasr2019comprehensive,shokri2017membership}. Although the adoption of DP during learning has been explored to mitigate this issue~\cite{abadi2016deep,wei2020federated}, the added noise can adversely affect model performance~\cite{yang2023privatefl}. Cryptographic alternatives, such as HE~\cite{fang2021privacy} and SMPC~\cite{mugunthan2019smpai} offer privacy without affecting models' performance; yet, such methods tend to prolong training times and increase communication and computation overhead. Recent efforts to optimize HE efficiency~\cite{mohammadi2023secure,zhang2020batchcrypt}, such as pruning or quantizing model updates prior to encryption, still struggle to maintain robust security with minimal computational demand on client devices. 

Recently, various FE schemes have been introduced for FL~\cite{qian2022cryptofe,xu2019hybridalpha,xu2022detrust}. These methods, offering lower cryptographic overhead and enabling complex computations like weighted averaging, present a significant improvement over HE approaches. Moreover, recent decentralized FE variations~\cite{chotard2018decentralized,chotard2020dynamic} address the privacy risks of HE's single-shared encryption key by using unique keys for each client. HybridAlpha~\cite{xu2019hybridalpha} and CryptoFE~\cite{qian2022cryptofe} utilize such decentralized FE schemes in FL, yet they require a fully trusted third-party entity, rendering them impractical in most FL scenarios. DeTrust-FL~\cite{xu2022detrust} overcomes this limitation by enabling clients a shared protocol between clients to collaboratively generate decryption keys; yet, imposing a fixed client participation agreement throughout the complete FL training process. Nevertheless, the aforementioned approaches directly apply the FE scheme in FL, encrypting all model parameters, which significantly increases their computation and communication overhead \textemdash~a major concern for devices with limited resources. Our work investigates efficient FE integration in FL, combining it with weight clustering and probabilistic filters, aiming to achieve both computational and communication efficiency while maintaining strong privacy.

\noindent \textbf{Communication-efficient FL.} Enhancing communication efficiency in FL is achievable through fast adaptation to downstream tasks by utilizing methods such as adaptive optimizers~\cite{reddi2020adaptive} or efficient client-sampling processes~\cite{chen2022optimal} that accelerate the convergence rate and, consequently minimize the data transmission requirement. Complementing these, model update compression techniques, such as sparsification~\cite{aji2017sparse,lin2017deep}, quantization~\cite{xu2021deepreduce,vargaftik2022eden,han2015deep_compression}, and low-rank approximation~\cite{mozaffari2021frl,mohtashami2022masked} can effectively reduce the data volume transmitted in each training round. Here, weight clustering~\cite{han2015deep_compression} offers a distinct opportunity by converting model parameters into discrete clusters, each represented by a single value. Unlike scalar quantization, which reduces the precision of each individual parameter, weight clustering limits only the number of unique weight values, facilitating efficient model update representations with minor performance degradation. Recent approaches that incorporate weight clustering in FL have achieved significant model compression rates~\cite{khalilian2023fedcode,tsouvalas2024communicationefficient}, substantially reducing communication costs during the training process, making them particularly useful for low-bandwidth communication networks.

In the realm of privacy-preserving FL, there is a growing interest in cryptographic schemes that need to operate in a communication-efficient manner with reduced computational overhead for clients~\cite{zhang2020batchcrypt,mohammadi2023secure}. SEFL~\cite{mohammadi2023secure} integrates Paillier HE with gradient pruning, effectively reducing the encrypted parameter count while minimally impacting model accuracy. BatchCrypt~\cite{zhang2020batchcrypt} optimizes the HE process by first quantizing and then encoding batches of gradients into a single long integer before encryption, allowing multiple model parameters to be encrypted in one operation, thus significantly easing the computational load on clients. However, BatchCrypt's efficiency is capped by the maximum number of values that can be encoded in a single integer, while it necessitates an adaption in the secure aggregation process for operating over the encoded batches. Our approach diverges existing methods by merging weight clustering with decentralized FE, surpassing the constraints of HE, while drastically increasing the number of parameters encrypted per operation and significantly decreasing the volume of data communicated in each training round. We further enhance privacy by transmitting the cluster-weights mapping through probabilistic filters, constructed via computationally inexpensive hashing operations. 

\section{Discussion\label{sec:discuss}}

In this work, we introduce ~\method, a framework aiming to bridge the gap between upholding privacy guarantees against inference attacks on model updates and delivering operational efficiency within FL. To achieve this, we opt to use model compression via weight clustering to transmit compressed representations of model updates during training, which we secure by combining decentralized FE with BF filter-based encoding. This integrated approach marks a significant stride towards developing FL systems where privacy and efficiency coexist as complementary rather than conflicting objectives.

\noindent \textbf{Limitations.} While we showed that~\method, when combined with DMCFE, can greatly improve clients' privacy with near-constant communication overhead and minimal impact on training times, its adaptation to other cryptographic frameworks for similar gains has not been explored within the scope of this study. We believe that integrating~\method~with emerging decentralized FE schemes, such as Dynamic Decentralized Functional Encryption (DDFE)\cite{chotard2020dynamic}, enabling clients to dynamically participate in the training process, or a Decentralized Trust Consensus (DTC) module~\cite{xu2022detrust}, eliminating the need for a TPA, could seamlessly integrate FE into existing FL systems and is a promising venue of future research. We also recognize the limitations imposed by pre-setting cluster sizes in weight clustering, which heavily relies on the complexity of both the model and the task, thereby limiting adaptability across diverse FL systems. Notably, the integration of recent adaptive weight clustering schemes~\cite{tsouvalas2024communicationefficient} within ~\method \textemdash~capable of dynamically adjusting cluster size based on model and task needs \textemdash~offers a promising path to overcome such challenges. \\

\noindent \textbf{Broader Impacts.} Our evaluation across widely utilized deep learning architectures in FL reveals a significant gap in recent research~\cite{xu2019hybridalpha, xu2022detrust, qian2022cryptofe, mohammadi2023secure, zhang2020batchcrypt}, emphasizing the need to assess performance under complex model architectures and challenging tasks. By open-sourcing our code, we aim to foster further exploration into approaches that simultaneously prioritize privacy and efficiency rather than treating them as separate challenges in FL. Furthermore, while~\method~did not consider eavesdropping threats, we highlight that the use of BF filter-based encoding, which depends on a unique seed value for accurate data reconstruction, naturally provides a safeguard against such threats. Additionally, in our evaluations, we primarily considered accuracy as a performance metric. However, as highlighted in~\cite{hooker2020characterising}, model compression techniques may disproportionately impact different subgroups of the data. We agree that this can create a fairness concern in all communication-efficient FL frameworks and deserves more attention from the community.

\section{Conclusions\label{sec:conclusions}}
\method~innovatively combines model compression via weight clustering, decentralized FE, and BF filter-based encoding to simultaneously tackle both challenges of privacy and efficiency in FL. Through rigorous evaluation across diverse datasets, architectures, and federated scenarios, ~\method~significantly lowers communication costs (>$13~\times$ reduction) and computational demands (>$4$-fold speedup) with minimal accuracy impact; thereby delivering robust privacy without reliance on trusted TPAs, especially for edge devices with limited computational and energy resources. Our framework not only offers a scalable solution for industries where data privacy and efficiency are paramount, but also paves the way for future advancements in secure, efficient, and privacy-preserving FL systems.

\section*{Acknowledgements}
\noindent The work presented in this paper is partially performed in the context of the Distributed Artificial Intelligent Systems (DAIS) project supported by the ECSEL Joint Undertaking (JU). JU receives support from the European Union's Horizon 2020 research and innovation programme and Sweden, Netherlands, Germany, Spain, Denmark, Norway, Portugal, Belgium, Slovenia, Czech Republic, Turkey.

\clearpage
\bibliographystyle{plainnat}
\bibliography{refs}

\newpage

\appendix

\section{Weight Clustering: Convergence \& Privacy Analysis}\label{asec:wc_privacy}

\noindent \textbf{\textit{Estimation Error Analysis due to Weight Clustering}}. In this section, we analyze the privacy implications due to the weight clustering process in \method. Recall that $\theta^{*}$ refers to the original post-trained model weights, while $\theta$ denotes the clustered weights. We encode cluster-weight mappings using probabilistic filters, which introduce an error probability of $2^{-\text{bpe}}$ (where $p$ denotes the false positive rate of the filter) leading to the assignment of a weight to an incorrect cluster. Note that the introduced error probability is independent across both clients and cluster dimensions. The estimation error between $\theta^{*}$ and $\theta$ can be computed as follows:

\begin{small}
    \begin{align}
            \mathbb{E} \left[ \left\| \theta^{*} - \theta \right\|_2^2 \right] &= \sum_{i=1}^{d} \mathbb{E} \left[ \left( \theta^{*}_{i} - \theta_{i} \right)^2 \right]  \\
            &= \sum_{i=1}^{d} \left( (1 - 2^{-\text{bpe}}) \cdot \mathbb{E} \left[ \left( \theta^{*}_{i} - c_{i} \right)^2 \right] + 2^{-\text{bpe}} \cdot \mathbb{E} \left[ \left( \theta^{*}_{i} - \tilde{c}_{i} \right)^2 \right] \right) \\
            &= \sum_{i=1}^{d} \left( (1 - 2^{-\text{bpe}}) \cdot \left( \sum_{k=1}^{\kappa} \sum_{\theta^{*}_{i} \in \mathcal{C}_k} \left\| \theta^{*}_{i} - c_k \right\|^2 \right) + 2^{-\text{bpe}} \cdot \left( \frac{1}{\kappa - 1} \sum_{k=1}^{\kappa} \left\| \theta^{*}_{i} - \tilde{c}_{i} \right\|^2 \right) \right).
    \end{align}
\end{small}

\noindent Here, $\tilde{c}_i$ refers to a randomly chosen centroid (any centroid apart from the correct one) due to the reconstruction error of the cluster-weights mapping. Assuming a uniform distribution of weights and centroids, the expected intra-cluster distance $\alpha$ (distance between weights within a given cluster) is given by $\alpha=\left( \sum_{k=1}^{\kappa} \sum_{\theta^{*}_{i} \in \mathcal{C}_k} \left\| \theta^{*}_{i} - c_k \right\|^2 \right)$, while the inter-cluster distance $\beta$ (error due to the false positive rate of the probabilistic filter) is estimated by the average distance from each given weight belonging to a cluster to all other clusters' centroids, computed as $\beta = \left( \frac{1}{\kappa - 1} \sum_{k=1}^{\kappa} \left\| \theta^{*}_{i} - \tilde{c}_{i} \right\|^2 \right)$. While exact estimation of $\alpha$ and $\beta$ are complex and depend on the specific characteristics of the data, we note that both $\alpha$ and $\beta$ are bounded. 

In terms of privacy amplifications due to weight clustering, we can consider the minimum discrepancy between $\theta^{*}$ and $\theta$ as worst case scenario, which occurs when both $\alpha$ and $\beta$ take their minimal values (referred by $D_{intra}$ and $D_{inter}$) across all $d$ weight values. Thus, we can derive the following:

\begin{align}\label{eq:wc_privacy}
    \mathbb{E} \left[ \left\| \theta^{*} - \theta \right\|_2^2 \right] &\ge d \left((1 - 2^{-\text{bpe}}) \cdot D_{intra} + 2^{-\text{bpe}} \cdot D_{inter} \right)
\end{align}

\noindent \textbf{\textit{Distributed Mean Estimation Error Analysis}}. We can now compute the the expected mean estimation error of server-side aggregated model to derive a privacy leakage estimation, similar to Equation~\ref{eq:wc_privacy}. For this, we compute the lower bound of the mean estimation error between the true mean is $\bar{\theta^{*}}^{r+1} = \frac{1}{N} \sum_{i=1}^{N} {\theta^{*}}_i^{r}$ and our estimation $\bar{\theta}^{r+1} = \frac{1}{N} \sum_{i=1}^{N} \theta_{i}^{r}$. Here, the mean estimation error is as follows:

\begin{small}
    \begin{align}\label{eq:wc_privacy_aggr}
    \mathbb{E}\left[ \left\| \bar{\theta^{*}}^{r+1} - \hat{\bar{\theta}}^{r+1} \right\|_2^2 \right]
    &= \sum_{i=1}^{d} \mathbb{E} \left[ \left ( \bar{\theta^{*}}^{r+1} - \bar{\theta}^{r+1} \right )^2 \right] \\
    &= \sum_{i=1}^{d} \mathbb{E} \left[ \left ( \frac{1}{N} \sum_{i \in \mathcal{N}} \left ( {\theta^{*}}_{i}^{r} - \theta_{i}^{r} \right ) \right )^2 \right] \\
    &= \frac{1}{N^2} \sum_{i=1}^{d} \mathbb{E} \left[ \left ( \sum_{i \in \mathcal{N}} \left ( {\theta^{*}}_{i}^{r} - \theta_{i}^{r} \right ) \right )^2 \right] \\
    &= \frac{1}{N^2} \sum_{i=1}^{d} \sum_{i \in \mathcal{N}} \mathbb{E} \left[ \left ( {\theta^{*}}_{i}^{r} - \theta_{i}^{r} \right )^2 \right] \\
    &\ge \frac{d \left((1 - 2^{-\text{bpe}}) \cdot \bar{D}_{intra} + 2^{-\text{bpe}} \cdot \bar{D}_{inter} \right)}{K}
    \end{align}
\end{small}

\noindent Here, $\bar{D}_{intra}$ and $\bar{D}_{inter}$ refer to the mean intra-cluster and inter-cluster distance across clients. Since server does not have access to clients' centroids, direct re-construction of clients' weights remains infeasible.

\end{document}